\documentclass[11pt,a4paper]{article}
\pdfoutput=1
\usepackage{jheppub}
\usepackage{latexsym,amsfonts,amsmath,amssymb}
\usepackage{bbm}
\usepackage{empheq}
\usepackage{graphicx}
\usepackage{color}
\usepackage{caption}
\usepackage{subcaption}
\usepackage[normalem]{ulem}
\usepackage{comment}
\usepackage{url}
\usepackage{slashed}
\usepackage{tabu}
\usepackage{multirow}
\usepackage{extarrows}
\usepackage{multicol}
\usepackage{amsmath}
\usepackage{mathtools}
\usepackage{float}
\usepackage{cancel}
\usepackage{amsxtra,graphics,epsfig,bm,tikz,xfrac,lscape}
\usepackage{tablefootnote}
\usetikzlibrary{decorations.pathmorphing}
\usetikzlibrary{decorations.markings}
\usetikzlibrary{arrows, decorations.markings, calc, fadings, decorations.pathreplacing, patterns, decorations.pathmorphing, positioning}
\usepackage{hyperref}
\usetikzlibrary{positioning, shapes}
\usetikzlibrary{chains}
\usetikzlibrary{arrows,fit,decorations.pathreplacing}
\tikzstyle{every picture}+=[remember picture]
\tikzstyle{na} = [baseline=-.5ex]

\newcommand{\NN}{\mathcal{N}}

\title{Dilaton Weyl multiplets for {$\mathcal{N}=3$} conformal supergravity in four dimensions}

\author{Soumya Adhikari$^{a}$,}
\author{Aravind Aikot$^{a}$,}
\author{Madhu Mishra$^{b}$ and}
\author{Bindusar Sahoo$^{a}$}

  \affiliation{$^a$ Indian Institute of Science Education and Research Thiruvananthapuram,
	Vithura, 695551, India  }
\affiliation{$^b$ Asia Pacific Center for Theoretical Physics, Postech, Pohang 37673, Republic of Korea }

\emailAdd{ssoumya.a012@gmail.com}
\emailAdd{arvd1719@alumni.iisertvm.ac.in}
\emailAdd{madhu.mishra@apctp.org}
\emailAdd{bsahoo@iisertvm.ac.in}

\usepackage{tikz}
\usetikzlibrary{shapes.geometric, arrows}

\tikzstyle{startstop} = [rectangle, rounded corners, 
minimum width=3cm, 
minimum height=1cm,
text centered, 
draw=black, 
fill=red!30]

\tikzstyle{io} = [trapezium, 
trapezium stretches=true, 
trapezium left angle=70, 
trapezium right angle=110, 
minimum width=3cm, 
minimum height=1cm, text centered, 
draw=black, fill=blue!30]

\tikzstyle{process} = [rectangle, 
minimum width=3cm, 
minimum height=1cm, 
text centered, 
text width=3cm, 
draw=black, 
fill=orange!30]

\tikzstyle{decision} = [diamond, 
minimum width=3cm, 
minimum height=1cm, 
text centered, 
draw=black, 
fill=green!30]
\tikzstyle{arrow} = [thick,->,>=stealth]

\abstract {We construct a dilaton Weyl multiplet for $\mathcal{N}=3$ conformal supergravity in four dimensions. We couple an on-shell vector multiplet to the standard Weyl multiplet and use the field equations of the vector multiplet to replace some of the components of the auxiliary fields of the standard Weyl multiplet with the fields of the vector multiplet and some dual gauge fields. The R-symmetry of the multiplet is $SU(2)\times U(1) \times U(1)$. Furthermore, we gauge fix one of the two $U(1)$ symmetries and rewrite the result for the dilaton Weyl multiplet with $SU(2)\times U(1)$ R-symmetry.}

\begin{document}

\maketitle

\bigskip

\section{Introduction}

Supergravity (SUGRA) represents a theoretical framework that synergistically integrates the principles of supersymmetry (SUSY) and general relativity (GR), thereby providing a more comprehensive and unified description of the universe. By incorporating gravity into the supersymmetric framework, SUGRA enables researchers to investigate the complicated links between gravitational forces and supersymmetric interactions. The implementation of supersymmetry in nature remains a topic of ongoing investigation. Nevertheless, it is evident that supersymmetry must be spontaneously broken, as the observed particle spectrum does not exhibit the characteristic supermultiplet structure with degenerate masses. 
 Moreover, the low-energy regime (100 GeV to 1 TeV) is theoretically described by  $\mathcal{N}=1$ supersymmetric theory (the number of supersymmetries is often denoted as $\mathcal{N}$), owing to its compatibility with chiral gauge theories inherent to the Standard Model \cite{Cremmer:1982en, Nilles:1983ge, Wess:1992cp, Martin:1997ns}. This framework suffers from an abundance of arbitrary parameters and representation choices. However, supersymmetric theories become increasingly constrained as the number of dimensions or supercharges increase. Consequently, it is desirable to derive the effective $\mathcal{N}=1$ model via spontaneous symmetry breaking from an extended supersymmetric theory ($\mathcal{N}>1$), thereby providing a more fundamental understanding of supersymmetry. By including gravity into these supersymmetric theories, we end up with a supergravity theory thereby combining gravity with other interactions in the standard model in a unified framework.
Although supergravity theories are not ultraviolet complete, they arise as an effective low energy description of an ultraviolet complete theory such as string theory in ten dimensions. This relationship again emphasizes SUGRA's significance in the larger context of unified theories. After compactifying a higher-dimensional theory, we may end up with a 4-dimensional theory with some amount of supersymmetry, based on the topology of the compactification manifold.\footnote{Compactification of 10D supergravity on various manifolds preserves supersymmetry as follows:
\begin{itemize}
    \item[a.] $T^6$ (6D torus): Full supersymmetry,
    \item[b.] $K3 \times T^2$: ($K3$ is a Calabi-Yau 2-fold): half supersymmetry,
    \item[c.]  Calabi-Yau 3-fold $(CY3)$ one-fourth of the original supersymmetry.
\end{itemize}}
The most basic and widely adopted compactification method is the Calabi-Yau compactification, which provides a simple and well-understood framework for compactifying extra dimensions. For example,
\begin{itemize}
    \item[1.] Starting with ten-dimensional heterotic supergravity, one can obtain $\mathcal{N}=4, D=4$ SUGRA when compactified on $T^6$, $\mathcal{N}=2, D=4$ SUGRA when compactified on $K3\times T^2$,  and $\mathcal{N}=1, D=4$ SUGRA when compactified on Calabi-Yau 3-fold ($CY3$).
\item[2.] Starting with ten-dimensional type IIB/IIA supergravity, one can obtain $\mathcal{N}=8, D=4$ SUGRA when compactified on $T^6$, $\mathcal{N}=4, D=4$ SUGRA when compactified on $K3\times T^2$, and $\mathcal{N}=2, D=4$ SUGRA, when compactified on $CY3$.    
\end{itemize}

\noindent
One distinguishing aspect of these supergravity theories is their large global symmetry groups and abelian gauge groups. Notably, none of the matter fields are charged under the abelian gauge group and hence such theories are called \textit{ungauged supergravity}. Furthermore, these theories provide a maximally supersymmetric Minkowski ground state in which all fields are massless. However, more complex compactifications such as sphere $S^6$ or compactification in the presence of flux, can lead to \textit{gauged supergravities}. Gauged supergravities can also be viewed as deformations of their ungauged counterparts, where non-vanishing background fluxes are incorporated by gauging specific global symmetries, resulting in the same type of gauged supergravity. These theories have non-abelian gauge symmetries, charged matter fields, and a scalar potential resulting from the internal geometry. 
One such compactification, particularly relevant to our study, is
\begin{itemize}
    \item[3.]compactification of 11D supergravity on a seven-dimensional tri-Sasakian manifold that yields minimal $\mathcal{N}=3$ gauged supergravity in four dimensions \cite{Termonia:1999cs,Cassani:2011fu, Karndumri:2016tpf}.
\end{itemize}
To bridge the gap between supersymmetric phenomenology and more fundamental theories, understanding matter coupling to extended supergravities in four dimensions is crucial. The most phenomenologically relevant models have $\mathcal{N}=2,\;3$, and $4$ supersymmetries, as $\mathcal{N}\geq 5$ supersymmetries cannot be coupled to matter multiplets, precluding viable theories with the $SU(3)\times SU(2) \times U(1)$ gauge group \cite{Haag:1974qh, Ferrara:1981qd}. It is noteworthy that scalar fields play a vital role in supersymmetry breaking, generating the scalar potential, determining vacuum expectation values, and inducing masses.


The formulation of extended supergravity theories is substantially advanced by the superconformal tensor calculus, a systematic and comprehensive framework that provides a robust methodology for constructing matter-coupled and higher-derivative supergravity theories. In this approach, the symmetry group of such supergravity theories, which is the super-Poincar{\'e} group, is leveraged by a larger symmetry group, specifically the superconformal group, thereby giving us a theory of conformal supergravity. The fundamental object in such a theory is the Weyl multiplet that encompasses the gauge fields associated with the superconformal group along with some covariant matter fields necessary to close the algebra off-shell i.e. without using the equations of motion. The introduction of covariant matter fields necessitates a modification of the supersymmetry transformation rules and the superconformal algebra, yielding a soft superconformal algebra characterized by structure functions dependent on the covariant matter fields \cite{Freedman:2012zz}. The off-shell nature of the Weyl multiplet plays a crucial role in giving us a systematic and tractable framework for the construction of the conformal supergravity theories, which is an intermediate step in the construction of higher derivative matter coupled Poincar{\'e} supergravity theories.


Based on the choice of the covariant matter fields, two distinct versions of the Weyl multiplet have been obtained in the literature, each with its unique characteristics and implications for supergravity theories. These multiplets go by the name of standard and dilaton Weyl multiplet. The standard Weyl multiplet is known in all dimensions between three and six and for various amounts of supersymmetries \cite{Ferrara:1977mv, deWit:1979dzm,Bergshoeff:1980is, Bergshoeff:1985mz,Fujita:2001kv,Butter:2013rba,Butter:2013goa}. The dilaton Weyl multiplet is also known for all such cases in dimensions greater than or equal to four \cite{Fujita:2001kv,Bergshoeff:1985mz,Bergshoeff:2001hc,Butter:2017pbp,Hutomo:2022hdi,Gold:2022bdk,Adhikari:2023tzi, Ciceri:2024xxf, Adhikari:2024esl}.\footnote{With the exception of $\mathcal{N}=2$ conformal supergravity in five dimensions where only dilaton Weyl multiplets are known \cite{Adhikari:2023tzi, Adhikari:2024esl}. There is no standard Weyl multiplet in this case because of the non-existence of a rigid superconformal algebra as per Nahm's classification \cite{Nahm:1977tg}.} The dilaton Weyl multiplet is distinguished by the inclusion of a dilaton field, which transforms non-trivially under the dilatation symmetry. This multiplet comprises fields that explicitly compensate for some of the additional symmetries present in conformal supergravity. Consequently, it is expected that the rigid limit of the soft superconformal algebra realized on the dilaton Weyl multiplet will reduce to a super-Poincaré algebra whereas the rigid limit of the soft superconformal algebra realized on the standard Weyl multiplet reduces to a rigid superconformal algebra \cite{Adhikari:2023tzi}.

In four spacetime dimensions, the standard Weyl multiplets for \( \mathcal{N}=2 \) and \( \mathcal{N}=4 \) supergravity theories were constructed in \cite{deWit:1979dzm} and \cite{Bergshoeff:1980is}, respectively. More recently, variants of the dilaton Weyl multiplets for \( \mathcal{N}=2 \) and \( \mathcal{N}=4 \) supergravity theories were obtained in \cite{Butter:2017pbp,Gold:2022bdk} and \cite{ Ciceri:2024xxf,Adhikari:2024esl}, respectively. Notably, the four-dimensional dilaton Weyl multiplet has been successfully employed to supersymmetrize arbitrary curvature squared invariants in \( \mathcal{N}=2 \) supergravity, as demonstrated in \cite{Mishra:2020jlc}.
The construction of the standard Weyl multiplet for  \( \mathcal{N}=3 \) supersymmetry had been a longstanding challenge due to its inherent complexity. Unlike the relatively tractable  \( \mathcal{N}=2 \) case and the highly constrained \( \mathcal{N}=4 \) scenario, the \( \mathcal{N}=3 \) supersymmetry presents a unique set of obstacles. Specifically, the unclear field content and supersymmetry algebra closure have hindered systematic investigations into  \( \mathcal{N}=3 \) conformal supergravity. However, recent advances \cite{vanMuiden:2017qsh, Hegde:2018mxv} have culminated in the successful construction of the standard Weyl multiplet for  \( \mathcal{N}=3 \), allowing the development of a full-fledged \( \mathcal{N}=3 \)
 conformal supergravity  in \cite{Hegde:2021rte} and Poincar\'e supergravity theory in \cite{Hegde:2022wnb}. Furthermore, the authors of \cite{Kuzenko:2023qkg} developed a superspace formulation of $\NN=3$ conformal supergravity in four dimensions.

\noindent
Motivated by the profound implications of scalar fields in phenomenological models, this paper undertakes a systematic and rigorous construction of the dilaton Weyl multiplet in four-dimensional $\mathcal{N}=3$ conformal supergravity. Our approach begins by coupling the standard Weyl multiplet to an on-shell vector multiplet \cite{Hegde:2022wnb}, which facilitates the incorporation of scalar degrees of freedom, enabling the derivation of the dilaton Weyl multiplet. The procedure involves the following steps:

\begin{enumerate}
	\item  The standard Weyl multiplet is coupled to an on-shell vector multiplet, resulting in a combined system that encodes the interactions between the gravitational and matter sectors.
	\item The field equations for the vector multiplet are reinterpreted as constraints on some of the covariant fields belonging to the standard Weyl multiplet, thereby allowing for solving them in terms of fields belonging to the vector multiplet along with some dual gauge fields.\footnote{The introduction of the dual gauge fields is necessary to solve some of the equations, as we will see later.}
   \item The remaining independent fields of the Weyl multiplet, along with the fields of the vector multiplet and the dual gauge fields, constitute the dilaton Weyl multiplet. 
\end{enumerate}

\noindent
The organization of the rest of the paper is as follows:

\noindent
In section \ref{N=3}, we provide a concise review of the standard Weyl multiplet and vector multiplet in the context of $\mathcal{N}=3$ conformal supergravity in four spacetime dimensions. This section also encompasses a detailed discussion of the equations of motion governing the vector multiplet, thereby establishing a foundational understanding of the underlying supersymmetric structure.
In section \ref{1dilaton4d}, we present a comprehensive and systematic construction of the $\mathcal{N}=3$ dilaton Weyl multiplet, where the breaking of the R-symmetry group $SU(3)\to SU(2)\times U(1) \times U(1)$ plays a crucial role. This section provides a detailed exposition of the theoretical framework underlying the dilaton Weyl multiplet, including the supersymmetry transformation of the component fields. In section \ref{2dilaton4d}, we extend our analysis to the $\mathcal{N}=3$ dilaton Weyl multiplet where the R-symmetry group is further reduced to $SU(2) \times U(1)$.
Finally, in section \ref{conclusions}, we summarize our findings, reflect on the implications of our results, and outline potential avenues for future research.

\section{{$\mathcal{N}=3$} Conformal Supergravity} \label{N=3}
An $\mathcal{N}$ extended conformal supergravity in four dimensions is based on the gauge theory of the superconformal algebra $SU(2,2|\mathcal{N})$,  satisfying certain curvature constraints. 
$\mathcal{N}=2$ and $\mathcal{N}=4$ conformal supergravities have been extensively studied in the past. However, detailed studies of $\mathcal{N}=3$ conformal supergravity began only in the last decade. The standard Weyl multiplet and the vector multiplet are the only known $\mathcal{N}=3$ multiplets currently. In the following subsections, we briefly review these multiplets.
 
\subsection{{$\mathcal{N}=3$}\label{stdweyl}  Standard Weyl Multiplet }
The $\mathcal{N}=3$ standard Weyl multiplet in four dimensions was constructed
in \cite{vanMuiden:2017qsh, Hegde:2018mxv}.
Table-\eqref{standard4d} gives the field content of the $\mathcal{N}=3$ standard Weyl multiplet in four dimensions.
\begin{table}[ht]
		\centering
		\centering
		\begin{tabular}{|c|c|c|c|c|}
			\hline
			Field&Properties&SU(3) irreps&$w$ & $c$\\
			\hline
			\multicolumn{5}{|c|}{Independent Gauge fields}\\
			\hline
			$e_\mu^a$&vielbein&\bf{1}&$-1$&$0$\\
			$\psi_\mu^I$&$\gamma_5 \psi_\mu^I=\psi_\mu^I$\;, gravitino &\bf{3}&$-1/2$&$-1/2$\\
			$V_\mu{}^{I}{}_J $&$(V_\mu{}^I{}_J)^*\equiv V_{\mu I}{}^J=-V_\mu{}^J{}_I $\;, $SU(3)_R$ gauge field &\bf{8}&0&0\\
			$A_\mu$&$ U(1)_R $ gauge field&\bf{1}&0&0\\
            $b_\mu$&Dilatation gauge field&\bf{1}&0&0\\
			\hline
			\multicolumn{5}{|c|}{Covariant fields}\\
			\hline
 $  T^I_{ab}$& Self-dual i.e $T^I_{ab}=\frac{1}{2}\varepsilon_{abcd}T^{I cd} $&\bf{3}&$1$&$1$\\
 $E_I$& Complex & $\bf{\Bar{3}}$ &$1$&$-1$\\
$ D^I{}_J$& $(D^I{}_J)^*\equiv D_I{}^J=D^J{}_I $&\bf{8}&2&0\\
$\chi_{IJ}$&$\gamma_5\chi_{IJ}=\chi_{IJ} $& $\bf{\Bar{6}}$&$3/2$&$-1/2$\\
$\zeta^I$ & $\gamma_5\zeta^I=\zeta^I $&\bf{3}& $3/2$&$-1/2$\\
$\Lambda_L$&$\gamma_5\Lambda_L=\Lambda_L$&\bf{1}&$1/2$&$-3/2$\\
 
			\hline
		\end{tabular}
		\caption{Field content of the $\mathcal{N}=3$ standard Weyl multiplet}
		\label{standard4d}	
	\end{table}

Here, Weyl weight ($w$) and chiral weight ($c$) tell us how the fields transform under dilatation and $U(1)$ R-symmetry, respectively. The Levi-Civita appearing in Table-\eqref{standard4d} as well at various other places in the paper is defined as $\varepsilon_{abcd}=i\epsilon_{abcd}$, where $\epsilon_{abcd}$ is the standard Levi-Civita with $\epsilon_{0123}=1$. Conformal supergravity comes with two kinds of supersymmetry transformations referred to as $Q$ (ordinary) supersymmetry and $S$ (special) supersymmetry. The $Q$ and $S$ supersymmetry transformations of the components of the standard Weyl multiplet parametrized by $\epsilon^I$ and $\eta^I$ respectively are given in equations-\eqref{standtrans}:
\allowdisplaybreaks
{
\begin{subequations}\label{standtrans}
\begin{align}
    \delta e_{\mu}^{a}&= \bar{\epsilon}_{I}\gamma^{a}\psi_{\mu}^{I}+\text{h.c.}\\[5pt]
	\delta \psi_{\mu}^{I}&=2\mathcal{D}_{\mu}\epsilon^{I}-\frac{1}{8}\varepsilon^{IJK}\gamma\cdot T_{J}\gamma_{\mu}\epsilon_{K}-\varepsilon^{IJK}\bar{\epsilon}_{J}\psi_{\mu K}\Lambda_{L}-\gamma_{\mu}\eta^{I} \label{gravitino}\\[5pt]
 \delta V_\mu{}^I{}_J &=\bar{\epsilon}^I\phi_{\mu J}- \frac{1}{48}\bar{\epsilon}^I\gamma_\mu\zeta_J+ \frac{1}{16}\varepsilon_{JKM}\bar{\epsilon}^K\gamma_\mu\chi^{IM}- \frac{1}{16}\bar{\epsilon}^I\gamma\cdot T_J \gamma_\mu\Lambda_R- \frac{1}{16}\bar{\epsilon}^I\gamma_\mu \Lambda_R E_J \nonumber \\
	&+\frac{1}{8}\varepsilon_{KMJ}E^I\bar{\epsilon}^K\psi_\mu^M  + \frac{1}{4}\bar{\epsilon}^I\gamma^a\psi_{\mu J}\bar{\Lambda}_L\gamma_a\Lambda_R-\bar{\psi}_\mu^I\eta_J-\text{h.c.}-\text{trace} \\[5pt]
 \delta A_\mu &=\frac{i}{6}\bar{\epsilon}^I\phi_{\mu I}+ \frac{i}{36}\bar{\epsilon}^I\gamma_\mu\zeta_I+ \frac{i}{12}\varepsilon_{KMP}E^P\bar{\epsilon}^K\psi_{\mu}^M+ \frac{i}{12}\bar{\epsilon}^I\gamma\cdot T_I\gamma_\mu\Lambda_R+\frac{i}{12}\bar{\epsilon}^I\gamma_\mu\Lambda_RE_I\nonumber\\
	&-\frac{i}{3}\bar{\epsilon}^I\gamma^a\psi_{\mu I}\bar{\Lambda}_L\gamma_a\Lambda_R-\frac{i}{6}\bar{\psi}_\mu^I\eta_I+\text{h.c.
}\\[5pt]
\delta b_\mu &= \frac{1}{2}(\bar{\epsilon}^I\phi_{\mu I}-\bar{\psi}_\mu^I\eta_I)+\text{h.c.}\\[5pt]
	\delta \Lambda_L&=-\frac{1}{4}E_I\epsilon^I+\frac{1}{4}\gamma\cdot T_I\epsilon^I\\[5pt]
	\delta E_I &=-4 \bar{\epsilon}_I\cancel{D}\Lambda_L-\frac{1}{2}\varepsilon_{IJK}\bar{\epsilon}^J\zeta^K+\frac{1}{2}\bar{\epsilon}^J\chi_{IJ}-\frac{1}{2}\varepsilon_{IJK}E^K\bar{\epsilon}^J\Lambda_L-4\bar{\Lambda}_L\Lambda_L\bar{\epsilon}_I\Lambda_R- 4\bar{\eta}_I\Lambda_L \\[5pt]
	\delta T^I_{ab} &= -\bar{\epsilon}^I\cancel{D}\gamma_{ab}\Lambda_R-4\varepsilon^{IJK}\bar{\epsilon}_JR_{ab}(Q)_K+\frac{1}{8}\bar{\epsilon}_J\gamma_{ab}\chi^{IJ}+\frac{1}{24}\varepsilon^{IJK}\bar{\epsilon}_J\gamma_{ab}\zeta_K \nonumber \\
	&-\frac{1}{8}\varepsilon^{IJK}E_J\bar{\epsilon}_K\gamma_{ab}\Lambda_R+\bar{\eta}^I\gamma_{ab}\Lambda_R \\[5pt]
	\delta \chi_{IJ}&=2\cancel{D}E_{(I}\epsilon_{J)}-8\varepsilon_{KM(I}\gamma\cdot R(V)^M{}_{J)}\epsilon^K-2\gamma\cdot\cancel{D}T_{(I}\epsilon_{J)}+\frac{1}{3}\varepsilon_{KM(I}D^M{}_{J)}\epsilon^K\nonumber \\
 &+\frac{1}{4}\varepsilon_{KM(I}E^K\gamma\cdot T_{J)}\epsilon^M-\frac{1}{3}\bar{\Lambda
	}_L\gamma_a\epsilon_{(I}\gamma^a\zeta_{J)}+\frac{1}{4}\varepsilon_{PM(I}E_{J)}E^M\epsilon^P-\bar{\Lambda}_L\gamma^a\Lambda_R\gamma_aE_{(I}\epsilon_{J)}\nonumber\\
	&-\bar{\Lambda}_L\gamma\cdot T_{(I}\gamma^a\Lambda_R\gamma_a\epsilon_{J)}+ 2\gamma\cdot T_{(I}\eta_{J)}+ 2E_{(I}\eta_{J)}\\[5pt]
	\delta \zeta^I &=- 3\varepsilon^{IJK}\cancel{D}E_J\epsilon_K +\varepsilon^{IJK}\gamma\cdot\cancel{D}T_K\epsilon_J-4\gamma\cdot R(V)^I{}_J\epsilon^J-16i\gamma\cdot R(A)\epsilon^I-\frac{1}{2}D^I{}_J\epsilon^J\nonumber\\
	&-\frac{3}{8}E^I\gamma\cdot T_J\epsilon^J+\frac{3}{8}E^J\gamma\cdot T_J\epsilon^I+\frac{3}{8}E^IE_J\epsilon^J+\frac{1}{8}E^JE_J\epsilon^I- 4 \bar{\Lambda}_L\cancel{D}\Lambda_{R}\epsilon^I- 4 \bar{\Lambda}_R\cancel{D}\Lambda_L\epsilon^I\nonumber\\
	&- 3\bar{\Lambda}_R\cancel{D}\gamma_{ab}\Lambda_L\gamma^{ab}\epsilon^I-3\bar{\Lambda}_L\gamma_{ab}\cancel{D}\Lambda_R\gamma^{ab}\epsilon^I+\frac{1}{2}\varepsilon^{IJK}\bar{\Lambda}_L\gamma^a\epsilon_J\gamma_a\zeta_K-6\bar{\Lambda}_L\Lambda_L\bar{\Lambda}_R\Lambda_R\epsilon^I\nonumber\\
	&+\varepsilon^{IJK}\gamma\cdot T_J\eta_K-3\varepsilon^{IJK}E_J\eta_K\\[5pt]
	\delta D^I{}_J&=-3\bar{\epsilon}^I\cancel{D}\zeta_J-3\varepsilon_{JKM}\bar{\epsilon}^K\cancel{D}\chi^{IM}+\frac{1}{4}\varepsilon_{JKM}\bar{\epsilon}^I\zeta^K E^M +\frac{1}{2}\varepsilon_{JKM}\bar{\epsilon}^K\zeta^M E^I +\frac{3}{4}\bar{\epsilon}^I\chi_{JK}E^K\nonumber\\
	&+ 3\bar{\epsilon}^I\gamma\cdot T_J\overset{\leftrightarrow}{\cancel{D}}\Lambda_R-3\bar{\epsilon}^I\cancel{D}\Lambda_RE_J-3\bar{\epsilon}^I\cancel{D}E_J\Lambda_R+ \frac{3}{4}\varepsilon_{JKM}E^M\bar{\epsilon}^K\Lambda_LE^I\nonumber\\
&+ {3\varepsilon_{JKM}T^I\cdot T^M\bar{\epsilon}^K\Lambda_{L}}-2\bar{\epsilon}^I\Lambda_L\bar{\Lambda}_R\zeta_J-3\bar{\epsilon}^I\Lambda_L\bar{\Lambda}_R\Lambda_RE_J+3\bar{\epsilon}^I\gamma\cdot T_J\Lambda_L\bar{\Lambda}_R\Lambda_R\nonumber\\
&+\text{h.c.} -\text{trace}
	\end{align}
    \end{subequations}
 }

 \textbf{Notation:} In four-dimensional conformal supergravity, chiral notation is followed throughout the existing literature, which we also follow here. In this notation, the raising and lowering of the R-symmetry indices is via complex conjugation. If the field is a real bosonic field or a Majorana fermion, this means that the raising and lowering of the R-symmetry indices would change the duality/chirality of the bosonic/fermionic field. For example, in $\mathcal{N}=3$ standard Weyl multiplet, we have a real anti-symmetric tensor field, whose self-dual component is given as $T_{ab}^{I}$. Under complex conjugation, it goes to $T_{abI}\equiv (T_{ab}^{I})^*$, which is the anti-self dual part of the same real anti-symmetric tensor field. In the same way, the left chiral part of a Majorana spinor denoted by $\zeta^I$, under an appropriate complex conjugation goes to the right chiral component $\zeta_I$ of the same Majorana spinor. The exact relation between them is given as $\zeta_I=i\gamma^0 C^{-1}(\zeta^I)^*$, where $C$ is the charge conjugation matrix. In $\mathcal{N}=3$ standard Weyl multiplet we also have some Majorana spinor which does not have any R-symmetry index. In that case, the left and right chiral parts are denoted as $\Lambda_L$ and $\Lambda_R$. But the relation between them is the same as above i.e. $\Lambda_R=i\gamma^0 C^{-1}(\Lambda_L)^*$. Note that the subscript $L$ and $R$ here, as well as elsewhere, should not be confused with the R-symmetry indices $I,J$, etc. 

 We also use $D_a$ to denote a fully supercovariant derivative. $\mathcal{D}_{\mu}\epsilon^I$ appearing in equation-\eqref{gravitino} is defined as:
 \begin{align}
     \mathcal{D}_{\mu}\epsilon^I=\partial_\mu \epsilon^I-\frac{1}{4}\gamma\cdot
      \omega_\mu \epsilon^I+\frac{1}{2}(b_\mu+iA_\mu)\epsilon^I-V_\mu{}^I{}_J \epsilon^J
 \end{align}

\subsection{{$\mathcal{N}=3$} Vector Multiplet} \label{vector}
The $\mathcal{N}=3$ vector multiplet coupled to the standard Weyl multiplet was constructed in \cite{Hegde:2022wnb}. The field contents of the vector multiplet are given in Table-\eqref{4dvector}. 
\begin{table}[h!]
	\centering
	\centering
	\begin{tabular}{ |c|c|c|c|c|c| }
		\hline
		Field & Type& SU(3)&$w$ (Weyl weight)&$c$ (chiral weight)\\
		\hline
		$C_\mu$& Boson&\bf{1}&0&0\\
		$\xi_I$&Boson&$\bar{\bf{3}}$&$1$&$-1$\\
		$\psi_{I}$&{Fermion}&$\bf{\bar{3}}$&$3/2$&$1/2$\\
  $\theta_L$ & Fermion & \bf{1}& $3/2$& $3/2$\\
		\hline
	\end{tabular}
	\caption{Field contents of the $\NN=3$ vector multiplet}
	\label{4dvector}	
\end{table}

The $Q$ and $S$ transformations of the components of this multiplet are given in \eqref{vec}.
\allowdisplaybreaks
{\begin{subequations}\label{vec}
\begin{align}
\delta{C}_{\mu}&= \bar{\epsilon}^{I}\gamma_{\mu}\psi_{I}-2\varepsilon_{IJK}\bar{\epsilon}^{I}\psi_{\mu}^J\xi^K  - \bar{\epsilon}_{I}\gamma_{\mu}\Lambda_L \xi^I +\text{h.c.}    \\
\delta \psi_{I}&= -\frac{1}{2}   \gamma ^{ab} \mathcal{F}^+_{ab} \epsilon_I - 2 \varepsilon_{IJK} \cancel{D} \xi^K \epsilon^J - \frac{1}{4} E_I \xi^J \epsilon_J + \frac{1}{2} \bar{\Lambda}_L \theta_L \epsilon_I\nonumber \\
& + \frac{1}{2}\gamma_a \epsilon^J \bar{\Lambda}_R \gamma^a \Lambda_L \xi^K \varepsilon_{IJK} + 2 \varepsilon_{IJK} \xi^J \eta^K 
   \\
\delta \theta_L &= - 2 \cancel{D} \xi^I \epsilon_I -  \gamma^a \bar{\Lambda}_L \gamma_a \Lambda_R \xi^I \epsilon_I + \frac{1}{4} \varepsilon_{IJK} E^I \xi^J \epsilon^K - \bar{\Lambda}_R \psi_I \epsilon^I  - 2 \xi^I \eta_I 
   \\
\delta \xi_I&=-\bar{\epsilon_I} \theta_R + \varepsilon_{IJK} \bar{\epsilon}^J \psi^K  
\end{align}    
\end{subequations}}

where $\mathcal{F}_{ab}^+$ is the self-dual part of  modified superconformal field strength of $C_\mu$ given as.
\begin{align}
 \mathcal{F}_{a b}^{+}= {F}_{a b}^{+}-\frac{1}{4} \bar{\Lambda}_R \gamma_{a b} \theta_R-\frac{1}{2} T_{a b}^I\xi_I
 \end{align}
where,
\begin{align}
   {F}_{ab} =2 e_{[a}^\mu e_{b]}^\nu \partial_\mu C_\nu - \left( \bar{\psi}^I_{[a} \gamma_{b]} \psi_I - \epsilon_{IJK} \bar{\psi}_a^I \psi_b^J \xi^K - \bar{\psi}_I{}_{[a} \gamma_{b]} \Lambda_L \xi^I + \text{h.c}\right) 
\end{align}
is the standard supercovariant field strength associated with $C_{\mu}$.
The field equations of the vector multiplet are given below in \eqref{eom},
\allowdisplaybreaks{
\begin{subequations} \label{eom}
    \begin{align}
& \cancel{D} \psi_I+\frac{1}{2} \bar{\Lambda}_R \psi_I \Lambda_L-\frac{1}{8} E_I \theta_L+\frac{1}{8} \gamma \cdot T_I \theta_L+\frac{1}{8} \chi_{I J} \xi^J+\frac{1}{24} \varepsilon_{IJK} \zeta^J \xi^K=0\;,\label{psi} \\[5pt]
& \cancel{D} \theta_R-\frac{3}{4} \bar{\Lambda}_R \theta_R \Lambda_L+\frac{1}{4} \gamma \cdot F^{-} \Lambda_L-\frac{3}{8} \bar{\Lambda}_L \Lambda_L \theta_L-\frac{1}{8} \gamma \cdot T_I \Lambda_L \xi^I-\frac{1}{8} E_I \psi^I-\frac{1}{8} \gamma \cdot T_I \psi^I \nonumber \\
& -\frac{1}{12} \zeta^I \xi_I-\frac{1}{8} E^I \xi_I \Lambda_L=0 \;,\label{theta}\\[5pt]
&\square_c \xi_J+ \frac{1}{4} D_a (\bar{\Lambda}_R \gamma_a \Lambda_L \xi_J ) + \frac{1}{4} \bar{\Lambda}_R \cancel{D} \xi_J \Lambda_L+\frac{1}{4} F^{-} \cdot T_J-\frac{1}{16} \bar{\Lambda}_L \gamma \cdot T_J \theta_L-\frac{1}{8} \xi^I T_I \cdot T_J\nonumber \\
&+\frac{1}{24} \bar{\zeta}_J \theta_R +\frac{1}{16} E_J \bar{\Lambda}_R \theta_R  -\frac{1}{16} \bar{\chi}_{M J} \psi^M-\frac{1}{48} \varepsilon_{J K L} \bar{\zeta}^L \psi^K-\frac{1}{48} D^L{ }_J \xi_L-\frac{1}{96} \xi_J E^K E_K\nonumber \\
&+\frac{1}{12} \xi_J\left(\bar{\Lambda}_R \cancel{D} \Lambda_L+\bar{\Lambda}_L \cancel{D} \Lambda_R\right)  +\frac{1}{12} \xi_J \bar{\Lambda}_R \Lambda_R \bar{\Lambda}_L \Lambda_L=0\;,\label{xi}\\[5pt]
&D_a\left({G}^{+a b}-{G}^{-a b}\right)=0\;,\label{maxwell}
\end{align}
\end{subequations}
}
where,
\begin{align}
{G}_{ab}^+=& - i{F}_{ab}^+ +\frac{i}{2} \bar{\Lambda}_R \gamma_{ab} \theta_R + iT_{ab}^I \xi_I
\end{align}
and $G^-_{ab}=(G^+_{ab})^*$. The superconformal de-Alembertian. is denoted by $\square_c\equiv D_{a}D^{a}$. 
\section{$SU(2)\times U(1) \times U(1)$ dilaton Weyl multiplet }\label{1st dilaton}
In \(\mathcal{N}=3\) conformal supergravity in four dimensions, there are no known dilaton Weyl multiplets. In this section, we give the details of the construction of a dilaton Weyl multiplet for \(\mathcal{N}=3\) conformal supergravity. In order to construct it, we first couple one \(\mathcal{N}=3\) vector multiplet to the \(\mathcal{N}=3\) standard Weyl multiplet and use the field equations \eqref{eom} of the vector multiplet to solve for some components in the standard Weyl multiplet in terms of the fields from the vector multiplet {along with} some dual gauge fields.\footnote{The dual gauge fields will arise, as we will see when we try to solve the Maxwell's equation and a scalar field equation.} The combination of the remaining fields of the standard Weyl multiplet, the fields of the vector multiplet and dual gauge fields will constitute the $\mathcal{N}=3$ dilaton Weyl multiplet. Although it is a mixture of the on-shell vector multiplet and the standard Weyl multiplet, it is still an off-shell multiplet, as the field equations of the vector multiplet have already been solved. 

In order to solve the vector multiplet field equations, we need to break the \(SU(3)\) R-symmetry to its subgroup \(SU(2)\times U(1)\). This ensures that some components of the standard Weyl multiplet auxiliary fields 
$\chi_{IJ}\;, \zeta_{I}$ and $D^I{}_J$ decouple from the remaining components in the field equations of the fermionic fields and the bosonic scalar field, respectively, and then we can use those field equations to algebraically solve them in terms of the vector multiplet fields. In order to break the R-symmetry, we first re-write all the components of the standard Weyl as well as vector multiplet in terms of $SU(2)\times U(1)$ representations. $SU(3)$ indices are denoted by $I$, taking values from 1 to 3. These indices decompose into $I=3$ and $SU(2)$ index: $i$, taking values from 1 to 2. We use the following representation of the $SU(3)$ invariant Levi-Civita  $\varepsilon_{IJK}$:
\begin{align}
\varepsilon_{ij3}&=\varepsilon_{ij}
\end{align} 
 The decomposition of the $SU(3)$ irreducible representations carried by the fields of the \(\mathcal{N}=3\) standard Weyl multiplet and the vector multiplet into \(SU(2)\times U(1)\) is given in appendix-\ref{decomposition} along with the reality conditions on the decomposed fields carrying \(SU(2) \times U(1)\) irreducible representations.

The vector multiplet field $\xi^I $ is instrumental in the symmetry breaking from $SU(3)$ to $SU(2) \times U(1)$. This field decomposes into $\xi^i$ and $\xi^3\equiv \xi $, as shown in appendix-\ref{decomposition}. We impose the gauge-fixing condition:
\begin{align}\label{gfcond}
     \xi^i=0.
\end{align}
to achieve the required symmetry breaking.

The Q-supersymmetry parameter $\epsilon^I$ decomposes into supersymmetry parameters carrying $SU(2)$ representation, $\epsilon^i$ and $\epsilon^3 \equiv \epsilon_L$.  The Q-supersymmetry transformation does not preserve the gauge condition \eqref{gfcond}. Hence, it needs to be redefined in the following way by adding a compensating $SU(3)$ transformation: 
\begin{align}
\delta_Q^{\text{new}}(\epsilon^i,\epsilon_L)= \delta_Q(\epsilon)+\delta_{SU(3)}(\Lambda^i{}_3=u(\epsilon)^i)   
\end{align}

 where $u(\epsilon)^i$  is a function of fields and the supersymmetry parameter (\(\epsilon\)) as given below:
\begin{align} \label{compensate}
    u(\epsilon)^i&=\frac{1}{{|\xi|}^2}\xi\bigg\{\bar{\epsilon}^i\theta_L -\varepsilon^{ij}\bigg(\bar{\epsilon}_j\psi_R-\bar{\epsilon}_R\psi_j \bigg)\bigg\}
\end{align}

Henceforth, in this section we will use $\delta_Q^{\text{new}}$ instead of $\delta_Q$ and therefore, we will drop the superscript ``new'' and refer to it as $\delta_Q$. 

The components of the $SU(3)$ gauge field $V_\mu{}^I{}_J$ are $V_\mu{}^i{}_j\;, V_\mu{}^i{}_3 $ and $V_\mu{}^3{}_3 $. Now let us decompose the $V_\mu{}^i{}_j $ component of the $SU(3)$ gauge field into the fields transforming in the $\bf{3}\oplus \bf{1}$ of $SU(2)$ as
\begin{align}
   V_\mu{}^i{}_j&\equiv v_\mu{}^i{}_j -\frac{i}{2}\delta^i{}_j\mathring{v}_\mu
\end{align}
 The field $v_\mu{}^i{}_j$ will play the role of the $SU(2)$ gauge field whereas $\mathring{v}_\mu\equiv -iV_\mu^3{}_3 $ will play the role of a new $U(1)$ gauge field. Note that we are denoting the $U(1)$ gauge field $\mathring{v}_{\mu}$ using a mathring since it will become a composite field in the dilaton Weyl multiplet. We will denote all composite fields using a mathring in this paper. We use $c_v$ and $c_A$ to denote the chiral weights of the fields w.r.t. the $U(1)$ transformations associated with the gauge fields $\mathring{v}_\mu$ and $A_\mu$, respectively. In Table-\eqref{1dilaton4d}, we give the values of $(c_v,c_A)$ for all the fields of the dilaton Weyl multiplet. Similarly, we will denote the $U(1)$ gauge transformation parameters associated with the gauge fields $\mathring{v}_\mu$ and $A_\mu$ as $\lambda_v$ and $\lambda_A$, respectively.

The component $V_a{}^i{}_3$ will turn into a covariant field after making appropriate modifications to it by adding some gravitino terms. The reason for making this modification is as follows. The new supersymmetry transformation involves $SU(3)$ compensating transformations with field-dependent parameters. When this compensating transformation acts on $V_a{}^i{}_3$, it gives a derivative of the supersymmetry parameters. To cancel the derivative of the parameter, we need to add explicit gravitino terms to $V_a{}^i{}_3$, which will turn it into a covariant field $Y_a^i$. The relation between them is given below in \eqref{Ya}: 
\begin{align}\label{Ya}
V_a{}^i{}_3-\frac{1}{2}u(\psi_a)^i&=Y^i_a    
\end{align}

The supercovariant derivatives $D_a$ that appear in the standard Weyl multiplet as well as in the vector multiplet are supercovariant w.r.t. the old supersymmetry transformations and the $SU(3)\times U(1)$ R-symmetry of the standard Weyl multiplet. However, for the dilaton Weyl multiplet, it will decompose into a new covariant derivative plus some covariant field terms containing $Y_a^i$ so that the new covariant derivative is covariant w.r.t. the new supersymmetry transformations as well $SU(2)\times U(1)\times U(1)$ R-symmetry of the dilaton Weyl multiplet. For example, the covariant derivative of the gaugino $\psi^i$ given in \eqref{vector} decomposes as follows, 

\begin{align}
    D_a^{\text{old}}\psi^i = D_a^{\text{new}} \psi^i - Y_a^i \psi_R
\end{align}

Only $D_a^{\text{new}}$ will appear in what follows, and we drop the superscript ``new" for convenience.
  
Now, let us rewrite the field equations \eqref{eom} in terms of the $SU(2)\times U(1)\times U(1)$ representations
as well as the new supercovariant derivative after imposing the gauge fixing condition \eqref{gfcond}:
\begin{subequations}\label{eom2}
\begin{align} 
&D_a\left(G^{+a b}-G^{-a b}\right)=0\;,\label{max1}\\[5pt]
    & \cancel{D} \psi_i - \slashed{Y}_{i} \psi_R +\frac{1}{2} \bar{\Lambda}_R \psi_i \Lambda_L-\frac{1}{8} E_i \theta_L+\frac{1}{8} \gamma \cdot T_i \theta_L+\frac{1}{8} \mathring{\chi}_{i} \bar{\xi}+\frac{1}{24} \varepsilon_{ij} \zeta^j \bar{\xi}=0\;,\label{psii} \\[5pt]
    & \cancel{D} \psi_R + \slashed{Y}^i \psi_i+\frac{1}{2} \bar{\Lambda}_R \psi_R \Lambda_L-\frac{1}{8} E \theta_L+\frac{1}{8} \gamma \cdot \mathring{T}^- \theta_L+\frac{1}{8}  \mathring{\chi}_L \bar{\xi}=0\;,\label{psir} \\[5pt]
    &\cancel{D} \theta_R-\frac{3}{4} \bar{\Lambda}_R \theta_R \Lambda_L+\frac{1}{4} \gamma \cdot F^{-} \Lambda_L-\frac{3}{8} \bar{\Lambda}_L \Lambda_L \theta_L-\frac{1}{8} \gamma \cdot \mathring{T}^- \Lambda_L \bar{\xi}-\frac{1}{8} E_i \psi^i -\frac{1}{8} E \psi_L -\frac{1}{8} \gamma \cdot T_i \psi^i \nonumber\\
    &-\frac{1}{8} \gamma \cdot \mathring{T}^- \psi_L  -\frac{1}{12} \mathring{\zeta}_L \xi - \frac{1}{8} \bar{E} \xi \Lambda_L = 0\;,\label{theta1} \\[5pt]
    &- D^a(Y_{ai} \xi) -Y_a^{i} D^a\xi +\frac{1}{2} \bar{\Lambda}_R \slashed{Y}_i \xi \Lambda_L+\frac{1}{4} F^{-} \cdot T_i-\frac{1}{16} \bar{\Lambda}_L \gamma \cdot T_i \theta_L-\frac{1}{8} \bar{\xi} \mathring{T}^-\cdot T_i+\frac{1}{24} \bar{\zeta}_i \theta_R\nonumber \\ 
    & +\frac{1}{16} E_i \bar{\Lambda}_R \theta_R -\frac{1}{16} \bar{\chi}_{ij} \psi^j -\frac{1}{16} \mathring{\bar{\chi}}_{i} \psi_L-\frac{1}{48} \varepsilon_{i j } \mathring{\bar{\zeta}}_L \psi^j + \frac{1}{48} \varepsilon_{ij} \bar{\zeta}^j \psi_L -\frac{1}{48} \mathring{D}_i \xi=0\;, \label{xii}\\[5pt]
    &D^aD_a \xi + Y^{ai}Y_{ai} \xi + \frac{1}{4} D_a (\bar{\Lambda}_R \gamma_a \Lambda_L \xi ) + \frac{1}{4} \bar{\Lambda}_R \cancel{D} \xi \Lambda_L+\frac{1}{4} F^{-} \cdot \mathring{T}^--\frac{1}{16} \bar{\Lambda}_L \gamma \cdot \mathring{T}^- \theta_L \nonumber \\
&-\frac{1}{8} \bar{\xi} \mathring{T}^- \cdot \mathring{T} +\frac{1}{24} \mathring{\bar{\zeta}}_R \theta_R+\frac{1}{16} E \bar{\Lambda}_R \theta_R-\frac{1}{16} \mathring{\bar{\chi}}_{j} \psi^j -\frac{1}{16} \mathring{\bar{\chi}}_L \psi_L-\frac{1}{48} \varepsilon_{ i j} \bar{\zeta}^j \psi^i-\frac{1}{48} \mathring{D} \xi\nonumber \\
&-\frac{1}{96} \xi E^k E_k -\frac{1}{96} \xi \lvert E \rvert ^2  +\frac{1}{12} \xi \left(\bar{\Lambda}_R \cancel{D} \Lambda_L+\bar{\Lambda}_L \cancel{D} \Lambda_R\right)+\frac{1}{12} \xi \bar{\Lambda}_R \Lambda_R \bar{\Lambda}_L \Lambda_L=0\;.\label{xi3} 
\end{align}
\end{subequations}

Maxwell's equation \eqref{max1} can be reinterpreted as a Bianchi identity if we think of $G_{ab}$ as the supercovariant field strength of an off-shell gauge field $\Tilde{C}_\mu$ which becomes a part of the multiplet. We can then use it to solve for $\mathring{T}^{\pm}_{ab}$ as shown in \eqref{Tab}, which makes it a composite field. 
\begin{align}\label{Tab}
    \mathring{T}_{ab}^+ &= \frac{1}{\xi} \bigg( {F}^+_{ab} - i {G}^+_{ab} -\frac{1}{2} \bar{\Lambda}_R \gamma_{ab} \theta_R \bigg) \nonumber \\
     \mathring{T}_{ab}^- &= ( \mathring{T}_{ab}^+ )^*
\end{align}
The introduction of a new off-shell gauge field $\Tilde{C}_{\mu}$ introduces a new $U(1)$ gauge symmetry associated with it. The corresponding parameter is referred to as $\Tilde{\lambda}$.

The field equation for $\psi_I$ \eqref{psi} decomposes into the field equations for $\psi_i $ and $\psi_R$. They are given in \eqref{psii} and \eqref{psir} respectively. It can be seen that the field $\chi_{ij}$ drops out from both the field equations as a result of the gauge condition \eqref{gfcond}. Similarly, the field $\mathring{\chi}_i$ and $\mathring{\chi}_L$ drops out from the field equation of $\psi_R$ and $\psi_i$ respectively. Hence, the field equations of $\psi_i$ and $\psi_R$ can be used to algebraically solve for the fields $\mathring{\chi}_i$ and $\mathring{\chi}_L$, respectively. We give the explicit expressions for $\mathring{\chi}_i$ and $\mathring{\chi}_L$ in \eqref{chisol} and \eqref{chiLsol},  respectively. The field $\chi_{ij}$ remains as an independent field.

It can be seen that the field $\zeta^i$ drops out in the field equation of $\theta_R$ due to the gauge fixing condition \eqref{gfcond}. The field $\mathring{\zeta}_L$ can be solved algebraically as shown in \eqref{zeta3sol}, and the field $\zeta^i$ becomes an independent field.
\begin{subequations}\label{fermsol}
\begin{align}
     \mathring{\chi}_i  & = -\frac{8}{ \bar{\xi}} \bigg(  \cancel{D} \psi_i - \slashed{Y}_{i} \psi_3 +\frac{1}{2} \bar{\Lambda}_R \psi_i \Lambda_L-\frac{1}{8} E_i \theta_L+\frac{1}{8} \gamma \cdot T_i \theta_L+\frac{1}{24} \varepsilon_{ij} \zeta^j \bar{\xi} \bigg)\;,\label{chisol} \\
    \mathring{\chi}_L  & = -\frac{8}{ \bar{\xi}} \bigg( \cancel{D} \psi_R + \slashed{Y}^i \psi_i+\frac{1}{2} \bar{\Lambda}_R \psi_R \Lambda_L-\frac{1}{8} E \theta_L+\frac{1}{8} \gamma \cdot \mathring{T}^- \theta_L \bigg)\;,\label{chiLsol} \\
    \mathring{\zeta}_L  & = \frac{12}{\xi}  \bigg(  \cancel{D} \theta_R-\frac{3}{4} \bar{\Lambda}_R \theta_R \Lambda_L+\frac{1}{4} \gamma \cdot \hat{F}^{-} \Lambda_L-\frac{3}{8} \bar{\Lambda}_L \Lambda_L \theta_L-\frac{1}{8} \gamma \cdot \mathring{T}^- \Lambda_L \bar{\xi}-\frac{1}{8} E_i \psi^i -\frac{1}{8} E \psi_L  \nonumber\\
    &-\frac{1}{8} \gamma \cdot T_i \psi^i-\frac{1}{8} \gamma \cdot \mathring{T}^- \psi_L   - \frac{1}{8} \bar{E} \xi \Lambda_L \bigg)\;.\label{zeta3sol}
\end{align}
\end{subequations}

The field equation for $\xi_I$ \eqref{xi} decomposes into the field equations for $\xi_i $ and $\xi$. They are given in \eqref{xii} and \eqref{xi3} respectively. It can be seen that the fields $D^i{}_j$ and $\mathring{D}$ drop out from the field equation of $\xi_i$ as a result of the gauge condition \eqref{gfcond}. Hence, the field $\mathring{D}_i$ can be solved algebraically using the field equation of $\xi_i$ and is completely determined in
terms of the other fields as shown in \eqref{Disol}. 

\begin{align}\label{Disol}
   \mathring{D}_i = &\frac{48}{\xi} \bigg( - D^a(Y_{ai} \xi) -Y_{ai} D_a \xi +\frac{1}{2} \bar{\Lambda}_R \slashed{Y}_i \xi \Lambda_L+\frac{1}{4} \widehat{F}^{-} \cdot T_i-\frac{1}{16} \bar{\Lambda}_L \gamma \cdot T_i \theta_L-\frac{1}{8} \bar{\xi} \mathring{T}^-\cdot T_i \nonumber \\
   & +\frac{1}{24} \bar{\zeta}_i \theta_R+\frac{1}{16} E_i \bar{\Lambda}_R \theta_R  -\frac{1}{16} \bar{\chi}_{ij} \psi^j -\frac{1}{16} \mathring{\bar{\chi}}_{i} \psi_L-\frac{1}{48} \varepsilon_{i j } \mathring{\bar{\zeta}}_L \psi^j + \frac{1}{48} \varepsilon_{ij} \bar{\zeta}^j \psi_L \bigg) 
\end{align}

Now, what remains is the field equation of $\xi$. One can see that due to the gauge condition \eqref{gfcond}, the field equation of $\xi$ only contains $\mathring{D}$ and the field $\mathring{D}_{i}$ drops out. We multiply the field equation of $\xi$ with $\bar{\xi}$ and break the resulting equation
into real and imaginary parts. We use the real part to solve for $\mathring{D}$ as shown below in \eqref{Dsol}.
\begin{align}\label{Dsol}
    &\mathring{D}=\frac{24}{{\lvert\xi \rvert}^2 } \bar{\xi}\bigg\{D^aD_a \xi + Y^{ai}Y_{ai} \xi + \frac{1}{4} D_a (\bar{\Lambda}_R \gamma_a \Lambda_L \xi ) + \frac{1}{4} \bar{\Lambda}_R \cancel{D} \xi \Lambda_L+\frac{1}{4} F^{-} \cdot \mathring{T}^-\theta_R\nonumber \\
& -\frac{1}{16} \bar{\Lambda}_L \gamma \cdot \mathring{T}^- \theta_L-\frac{1}{8} \bar{\xi} \mathring{T}^- \cdot \mathring{T}+\frac{1}{24} \mathring{\bar{\zeta}}_R \theta_R +\frac{1}{16} E \bar{\Lambda}_R \theta_R  -\frac{1}{16} \mathring{\bar{\chi}}_{j} \psi^j -\frac{1}{16} \mathring{\bar{\chi}}_L \psi_L-\frac{1}{48} \varepsilon_{ i j} \bar{\zeta}^j \psi^i \nonumber \\
&-\frac{1}{96} \xi E^k E_k -\frac{1}{96} \xi \lvert E \rvert ^2+\frac{1}{12} \xi \left(\bar{\Lambda}_R \cancel{D} \Lambda_L+\bar{\Lambda}_L \cancel{D} \Lambda_R\right)  +\frac{1}{12} \xi \bar{\Lambda}_R \Lambda_R \bar{\Lambda}_L \Lambda_L +\text{h.c.}\bigg\}
\end{align}

The imaginary part of the field equation for $\xi$ can be rewritten as
\begin{align}
& D^a \bigg(\bar{\xi} D_a \xi - \xi D_a \bar{\xi} + \frac{1}{2} \bar{\theta}_R \gamma_a \theta_L + \bar{\psi}^j \gamma_a \psi_j + \frac{1}{2} \bar{\psi}_L \gamma_a \psi_R + \frac{1}{2} \xi \bar{\xi} \bar{\Lambda}_R \gamma_a \Lambda_L \bigg) =\frac{1}{8}F\cdot \Tilde{F}+\frac{1}{8}G\cdot \Tilde{G}  
\end{align}
where $\Tilde{F}=F^+-F^-$ and $\tilde{G}=G^+-G^-$.\\
The above equation can be interpreted as the Bianchi identity of a 3-form supercovariant field $H_{abc}$ corresponding to a two-form gauge field $B_{\mu \nu}$ as given below in \eqref{bianchi}.
\begin{align} \label{bianchi}
    D_{[a}H_{bcd]} &= -\frac{3}{8} ({F}_{[ab}F_{cd]}+{G}_{[ab}G_{cd]}) 
\end{align}
where we define the three-form field strength $H_{abc}$ as the dual of a one-form as given below in \eqref{defH}.
\begin{align} \label{defH}
    \frac{1}{3!} \varepsilon_{abcd} H^{bcd} &= \bar{\xi} D_a \xi - \xi D_a \bar{\xi} + \frac{1}{2} \bar{\theta}_R \gamma_a \theta_L + \frac{1}{2} \bar{\psi}^j \gamma_a \psi_j + \frac{1}{2} \bar{\psi}_L \gamma_a \psi_R + \frac{1}{2} \xi \bar{\xi} \bar{\Lambda}_R \gamma_a \Lambda_L 
\end{align}
Expanding the covariant derivatives in the R.H.S of \eqref{defH} gives:
\begin{align}\label{U1dependence}
 \frac{1}{3!} \varepsilon_{abcd} H^{bcd} &= \bar{\xi}\partial_a\xi-\xi\partial_a\bar{\xi}+2iA_a\bar{\xi}\xi+2i\mathring{v}_a\bar{\xi}\xi +(\frac{1}{2}\bar{\xi}\bar{\psi}_{a,R}\theta_R-\frac{1}{2}\bar{\xi}\bar{\psi}_a^j\psi^k\varepsilon_{jk} -h.c ) \nonumber\\
 &+\frac{1}{2}\bar{\theta}_R\gamma_a\theta_L+\frac{1}{2}\bar{\psi}^j\gamma_a\psi_j+\frac{1}{2}\bar{\psi}_L\gamma_a\psi_R + \frac{1}{2} \xi \bar{\xi} \bar{\Lambda}_R \gamma_a \Lambda_L 
\end{align}
From the above equation in \eqref{U1dependence}, we can solve for the $U(1)$ gauge field $v_\mu$ in terms of the three form field strength $H_{abc}$ and other fields as given below in \eqref{vasol}:\footnote{We have a choice of either making $A_\mu$ or $\mathring{v}_\mu$ gauge fields composite.} , 
\begin{align}\label{vasol}
 \mathring{v}_a =& \frac{i}{2\xi\bar{\xi}} \bigg(- \frac{1}{3!} \varepsilon_{abcd} H^{bcd} + \bar{\xi}\partial_a\xi-\xi\partial_a\bar{\xi}+2iA_a\bar{\xi}\xi +(\frac{1}{2}\bar{\xi}\bar{\psi}_{a,R}\theta_R-\frac{1}{2}\bar{\xi}\bar{\psi}_a^j\psi^k\varepsilon_{jk} -h.c ) \nonumber\\
 &+\frac{1}{2}\bar{\theta}_R\gamma_a\theta_L+\frac{1}{2}\bar{\psi}^j\gamma_a\psi_j+\frac{1}{2}\bar{\psi}_L\gamma_a\psi_R + \frac{1}{2} \xi \bar{\xi} \bar{\Lambda}_R \gamma_a \Lambda_L \bigg)
\end{align}
From this definition, we find the supersymmetry transformations of the 3-form field strength which are given as:
\begin{align}
    \delta_Q \bigg( \frac{1}{3!} \varepsilon_{abcd}H^{bcd} \bigg) & = \bar{\epsilon}_R \gamma_{ab} D^b (\theta_R \bar{\xi} ) - \varepsilon_{ij} \bar{\epsilon}^i \gamma_{ab} D^b (\psi^j \bar{\xi}) +\frac{1}{16} \varepsilon_{ij} \bar{\xi}  \bar{\theta}_R \gamma \cdot T^i \gamma_a \epsilon^j + \bar{\xi} \bar{\theta}_R \gamma_{ab} E^{bj} \epsilon_j \nonumber \\ & - \frac{1}{16} \bar{\xi} \bar{\psi}^i \gamma \cdot T_i \gamma_a \epsilon_R + \frac{1}{16} \bar{\xi} \bar{\psi}^i \gamma \cdot \mathring{T}^- \gamma_a \epsilon_i + \varepsilon_{ij} \bar{\xi} \bar{\epsilon}_L \gamma_{ab} \psi^i E^{bj} \nonumber \\
 & - \frac{1}{8} \bar{\epsilon} _i \gamma_a \gamma \cdot (F^- + iG^-)  \psi_i -\frac{1}{8} \bar{\epsilon}_i \gamma \cdot (F^+ + i G^+) \gamma_a \psi_i \nonumber \\
 & -\frac{1}{8} \bar{\epsilon}_R \gamma_a \gamma \cdot (F^- + i G^- )  \psi_L - \frac{1}{8} \bar{\epsilon}_R \gamma \cdot (F^+ + i G^+ ) \gamma_a \psi_L \nonumber \\
    & + \frac{1}{8}\bar{\xi} \bar{\epsilon}_R \gamma_a \gamma \cdot (F^- - i G^-) \Lambda_L + \frac{1}{8} \bar{\xi} \bar{\epsilon}_R \gamma \cdot (F^+ - iG^+) \gamma_a \Lambda_L  \\
\delta_S\bigg( \frac{1}{3!} \varepsilon_{abcd}H^{bcd} \bigg)&= -\frac{3}{2} \bar{\xi} \bar{\theta}_R \gamma_a \eta_L -\frac{3}{2} \varepsilon_{jk} \bar{\psi}^j \gamma_a \eta^k
\end{align}

The Bianchi identity \eqref{bianchi}, the supersymmetry transformation above, and the knowledge of superconformal soft algebra suffice to find the full transformations of the two form gauge field $B_{\mu \nu}.$  
\begin{align}
    \delta_Q B_{\mu \nu} &= 2 \partial_{[\mu} \Lambda_{\nu ]}- \frac{\lambda}{4} F_{\mu \nu} - \frac{\Tilde{\lambda}}{4} G_{\mu \nu}+\left(\bar{\epsilon}_R \gamma_{\mu \nu} \theta_R \bar{\xi} + \varepsilon_{ij} \bar{\epsilon}^i \gamma_{\mu \nu} \psi^j \bar{\xi} - 2 \xi \bar{\xi} \bar{\epsilon}_R \gamma_{[\mu} \psi_{\nu] L} \right.\nonumber \\
    &\left. + 2 \xi \bar{\xi} \bar{\epsilon}^i \gamma_{[\mu} \psi_{\nu ]i} + \text{h.c} \right) + \frac{1}{2} C_{[\mu} \delta C_{\nu]}  + \frac{1}{2} \Tilde{C}_{[\mu} \delta  \Tilde{C}_{\nu]} 
\end{align}
where $\Lambda_\mu$ is the gauge transformation parameter associated with the two-form gauge field $B_{\mu\nu
}$ and $\lambda\;,\tilde{\lambda}$ are the gauge transformation parameters associated with the $U(1)$ gauge fields $C_\mu$ and $\tilde{C}_\mu$ respectively. 

From the above transformation of $B_{\mu\nu}$, one can obtain the explicit form of the fully supercovariant three-form field strength $H_{\mu\nu\rho}$ given below in \eqref{Habc}: 
\begin{align}\label{Habc}
    H_{\mu \nu \rho} & = 3 \partial_{[\mu} B_{\nu \rho] } + \frac{3}{4} C_{[\mu} F_{\nu \rho ]} + \frac{3}{4} \Tilde{C}_{[\mu} G_{\nu \rho] } +\left(- \frac{3}{2} \bar{\psi}_{R [\mu} \gamma_{\nu \rho]} \theta_R -\frac{3}{2} \varepsilon_{ij} \bar{\psi}^i_{[\mu} \gamma_{\nu \rho]} \psi^j \bar{\xi}\right. \nonumber \\
    & \left.+ 3 \xi \bar{\xi} \psi_{R [\mu} \gamma_{\nu} \psi_{\rho] L} - 3 \xi \bar{\xi} \psi_{[\mu}^i \gamma_\nu \psi_{\rho] i} + \text{h.c}\right) 
\end{align}

Now, we summarise the results. The field equations \eqref{eom2} make some
fields of the standard Weyl multiplet composite. The remaining fields of the standard Weyl multiplet, together
with the fields of the vector multiplet and some dual gauge fields constitute the dilaton Weyl
multiplet of $\mathcal{N} = 3 $ conformal supergravity where the R-symmetry is $ SU(2)\times U(1)\times U(1)$. In Table-\eqref{1dilaton4d}, we give full details of the independent fields of the dilaton Weyl multiplet. In equations-\eqref{1dilatontrans} below, we give
the $Q$ and $S$ transformations of the independent fields of the dilaton Weyl multiplet.
\begin{table}[ht]
		\centering
		\centering
		\begin{tabular}{|c|c|c|c| c |}
			\hline
			Field&Properties&SU(2) Irreps& $w$&$(c_A, c_v)$\\            
			\hline
			\multicolumn{5}{|c|}{Independent Gauge fields}\\
			\hline
			$e_\mu^a$&vielbein&\bf{1}&$-1$&$(0,0)$\\
			$\psi_\mu^i$& $\gamma_5 \psi_\mu^i=\psi_\mu^i$,  &\bf{2}&$-1/2$&$(-1/2, -1/2)$\\
            $\psi_{\mu L}$& $\gamma_5 \psi_{\mu L}=\psi_{\mu L}$,  &\bf{2}&$-1/2$&$(-1/2, 1)$\\
			$v_\mu^{i}{}_j $&$(v_\mu^i{}_j)^*\equiv v_{\mu i}{}^j=-v_\mu{}^j{}_i $&\bf{3}&$0$&$(0,0)$\\
   			$A_\mu$&$ U(1)_A $ gauge field&\bf{1}&$0$&$(0,0)$\\
      $b_\mu$&Dilatation gauge field&\bf{1}&$0$&$(0,0)$\\
      $B_{\mu \nu}$& Two-form gauge field& \bf{1}&$0$&$(0,0)$\\
            $C_\mu$ & $ U(1) $ gauge field&\bf{1}&0&(0,0)\\
              $\Tilde{C}_\mu$ & $ U(1) $ gauge field&\bf{1}&0&(0,0)\\
			\hline
			\multicolumn{5}{|c|}{Covariant fields}\\
			\hline
   $Y_a^i$&Boson&\bf{2}&1&(0, $-\frac{1}{2}$)\\
 $  T_{ab}^i$& $T^i_{ab}=\frac{1}{2}\varepsilon_{abcd}T^{icd} $&\bf{1}&1&(1,$-1/2)$\\
 $E_i$& Complex & $\bf{\Bar{2}}$ &1&(-1, $1/2)$\\
  $E$& Complex & $\bf{1}$ &1&(-1,-1 )\\
$D^i{}_j$& $(D^i{}_j)^*\equiv D_i{}^j=D^j{}_i $, $D^i{}_i=0$& \bf{3}&2&(0,0)\\ 
$\chi_{ij} $&$\gamma_5\chi_{ij} =\chi_{ij} $& $\bf{\Bar{3}}$&${3}/{2}$&$(-1/2, 1)$\\
$\Lambda_L$&$\gamma_5\Lambda_L=\Lambda_L$&\bf{1}&${1}/{2}$&$(-3/2,0)$\\
$\zeta^i$& $\gamma_5\zeta^i=\zeta^i$&\bf{2}& ${3}/{2}$&$(-1/2, -1/2)$\\
$\xi $& Boson, Complex field & \bf{1} & $1$& $(-1,-1)$\\
$\psi_i$&$\gamma_5 \psi_i = - \psi_i$&\bf{2}&$3/2$&$(1/2, 1/2)$\\
$\psi_R$& $\gamma_5 \psi_R = - \psi_R$& \bf{1}&$3/2$&$(1/2,-1) $\\
$\theta_L $& $\gamma_5 \theta_L = \theta_L$ & \bf{1}&$3/2$&($3/2$,0)\\
 			\hline
		\end{tabular}
		\caption{Field content of the $\mathcal{N}=3$ dilaton Weyl multiplet with R-symmetry $SU(2)\times U(1)\times U(1)$}
		\label{1dilaton4d}	
	\end{table}
\allowdisplaybreaks
{\begin{subequations}\label{1dilatontrans}
\begin{align}
\delta e_\mu^a&=\bar{\epsilon}_i\gamma^a\psi_\mu^i+\bar{\epsilon}_R\gamma^a\psi_{\mu L}+\text{h.c.} \\[5pt]
\delta\psi_\mu^i&=2\mathcal{D}_\mu\epsilon^i-2Y^i_\mu\epsilon_L-u(\psi_\mu)^i\epsilon_L-\frac{1}{8} \varepsilon^{ij} \gamma\cdot \bigg(T_j\gamma_\mu\epsilon_R- \mathring{T}^-\gamma_\mu\epsilon_j \bigg)\nonumber\\
    &-\varepsilon^{ij}\bar{\epsilon}_j\psi_{\mu R}\Lambda_L +\varepsilon^{ij}\bar{\epsilon}_R \psi_{\mu j}\Lambda_L+ u(\epsilon)^i\psi_{\mu L}-\gamma_\mu \eta^i \\[5pt]
     \delta \psi_{\mu L}&=2\mathcal{D}_\mu\epsilon_L+2Y_{\mu j}\epsilon^j + u(\psi_\mu)_i \epsilon^i -\frac{1}{8}\varepsilon^{ij}\gamma\cdot T_i\gamma_\mu \epsilon_j-\varepsilon^{ij}\bar{\epsilon}_i\psi_{\mu j}\Lambda_L-u(\epsilon)_i\psi^i_\mu \nonumber\\
     &-\gamma_\mu\eta_R   \\
     \delta v_\mu{}^i{}_j&= \bar{\epsilon}^i\phi_{\mu j} -\frac{1}{48} \bar{\epsilon}^i \gamma_\mu \zeta_j +\frac{1}{16}\bigg(\varepsilon_{jk}\bar{\epsilon}^k\gamma_\mu \mathring{\chi}^i-\varepsilon_{jk}\bar{\epsilon}_L\gamma_\mu\chi^{ik} \bigg)\nonumber\\
    &-\frac{1}{16}\bar{\epsilon}^i\gamma\cdot T_j\gamma_\mu\Lambda_R-\frac{1}{16}\bar{\epsilon}^i\gamma_\mu\Lambda_RE_j+\frac{1}{4}\bar{\epsilon}^i\gamma^a\psi_{\mu j}\bar{\Lambda}_L\gamma_a\Lambda_R +\frac{1}{8}E^i\varepsilon_{jk}\bigg(\bar{\epsilon}^k\psi_{\mu L}-\bar{\epsilon}_L\psi^k_\mu \bigg) \nonumber\\
    &-u(\epsilon)^i\bigg(Y_{\mu j}+\frac{1}{2} u(\psi_\mu)_j \bigg)-\bar{\psi}_\mu^i\eta_j-\text{h.c.}-\text{trace}\\[5pt]
    \delta b_\mu&=\frac{1}{2}\bigg(\bar{\epsilon}^i\phi_{\mu i}+\bar{\epsilon}_L\phi_{\mu L}-\bar{\psi}_\mu^i\eta_i-\bar{\psi}_{\mu L}\eta_L \bigg)+\text{h.c.} \\[5pt]
    \delta A_\mu&= \frac{i}{6}\bigg( \bar{\epsilon}^i\phi_{\mu i}+\bar{\epsilon}_L\phi_{\mu L}\bigg)+\frac{i}{36}\bigg( \bar{\epsilon}^i\gamma_\mu\zeta_i+\bar{\epsilon}_L\gamma_\mu \mathring{\zeta}_R\bigg)+\frac{i}{12}\varepsilon_{kl}\bigg(\bar{E}\bar{\epsilon}^k\psi^l_\mu-E^l\bar{\epsilon}^k\psi_{\mu L}\nonumber\\
    &+E^l\bar{\epsilon}_L\psi^k_\mu
    \bigg) +\frac{i}{12}\bigg(\bar{\epsilon}^i\gamma\cdot T_i\gamma_\mu\Lambda_R+\bar{\epsilon}_L\gamma\cdot T^-\gamma_\mu \Lambda_R \bigg)+\frac{i}{12}\bigg(E_i\bar{\epsilon}^i\gamma_\mu\Lambda_R+E\bar{\epsilon}_L\gamma_\mu \Lambda_R \bigg) \nonumber\\
    &-\frac{i}{3}\bigg( \bar{\epsilon}^i\gamma^a\psi_{\mu i}+\bar{\epsilon}_L\gamma_a\psi_{\mu R}\bigg)\bar{\Lambda}_L\gamma_a\Lambda_R-\frac{i}{16}\bigg(\bar{\psi}_\mu^i\eta_i+\psi_{\mu L}\eta_L \bigg)+\text{h.c.}\\[5pt]
    \delta{C}_{\mu}&= \bar{\epsilon}^{i}\gamma_{\mu}\psi_{i}+ \bar{\epsilon}_{L}\gamma_{\mu}\psi_{R} -2\bar{\epsilon}^{i}\psi_{\mu}^j \bar{\xi} \varepsilon_{ij} - \bar{\epsilon}_{R}\gamma_{\mu}\Lambda_L \bar{\xi} +\text{h.c.} \\[5pt]
    \delta\Tilde{C}_\mu&= -i\bar{\epsilon}^i\gamma_\mu\psi_i-i\bar{\epsilon}_L\gamma_\mu \psi_R+2i\bar{\epsilon}^i\psi_\mu^j\varepsilon_{ij}\bar{\xi}+i\bar{\epsilon}_R\gamma_\mu\Lambda_L\bar{\xi} +\text{h.c.} \\[5pt]
   \delta B_{\mu \nu} &= (\bar{\epsilon}_R \gamma_{\mu \nu} \theta_R \bar{\xi} + \varepsilon_{ij} \bar{\epsilon}^i \gamma_{\mu \nu} \psi^j \bar{\xi} - 2 \xi \bar{\xi} \bar{\epsilon}_R \gamma_{[\mu} \psi_{\nu] L} + 2 \xi \bar{\xi} \bar{\epsilon}^i \gamma_{[\mu} \psi_{\nu ]i}+\text{h.c} )\nonumber\\
   & +\frac{1}{2}\left(C_{[\mu}\delta C_{\nu]}+\tilde{C}_{[\mu}\delta C_{\nu]}\right) \;,\label{Bmunu}\\[5pt]
    \delta Y_a^i& = \bar{\epsilon}^i D_a(\frac{\theta_L}{\bar{\xi}})- \varepsilon^{ij} \bar{\epsilon}_j D_a(\frac{\psi_R}{\bar{\xi}})+ \varepsilon^{ij} \bar{\epsilon}_R D_a(\frac{\psi_j}{\bar{\xi}}) -\frac{1}{48} \bar{\epsilon}^i \gamma_a \mathring{\zeta}_R + \frac{1}{48} \bar{\epsilon}_R \gamma_a \zeta^i \nonumber \\
    &  + \frac{1}{16} \varepsilon_{jk} \bar{\epsilon}^j \gamma_a \chi^{ik}- \frac{1}{16} \varepsilon^{ij} \bar{\epsilon}_j \gamma_a \mathring{\chi}_L + \frac{1}{16} \varepsilon^{ij} \bar{\epsilon}_R \gamma_a \mathring{\chi}_j - \frac{1}{16} \bar{\epsilon}^i \gamma \cdot \mathring{T}^- \gamma_a \Lambda_R  \nonumber \\
    &+ \frac{1}{16} \bar{\epsilon}_R \gamma \cdot T^i \gamma_a \Lambda_L - \frac{1}{16} \bar{\epsilon}^i \gamma_a \Lambda_R E + \frac{1}{16} \bar{\epsilon}_R \gamma_a \Lambda_L E^i - \frac{1}{2 \bar{\xi}} \bar{\theta}_L \bigg( -2Y_a^i \epsilon_L\nonumber \\
    &-\frac{1}{8} \varepsilon^{ij}  \gamma \cdot T_j \gamma_a \epsilon_R + \frac{1}{8} \varepsilon^{ij}  \gamma \cdot \mathring{T}^- \gamma_a \epsilon_j \nonumber \bigg) -\frac{1}{\bar{\xi}} \varepsilon^{ij}Y_{aj}\bar{\psi}_R \epsilon_R  + \frac{1}{2\bar{\xi}} \bar{\psi}_R  \bigg( \frac{1}{8} \gamma \cdot T^i \gamma_a \epsilon_L \nonumber \\
    &- \frac{1}{8} \gamma \cdot \mathring{T}^+ \gamma_a \epsilon^i \bigg) + \frac{1}{8 \bar{\xi}} \bar{\psi}_j \gamma \cdot T^{[i} \gamma_a \epsilon^{j]} -\frac{1}{\bar{\xi}} \varepsilon^{ij} \bar{\psi}_j \epsilon_k Y_a^k+ \frac{1}{2 \bar{\xi}} \bar{\theta}_L \gamma_a \eta^i - \frac{1}{2\bar{\xi}} \varepsilon^{ij} \bar{\psi}_R \gamma_a \eta_j \nonumber\\
    &+ \frac{1}{2\bar{\xi}} \epsilon^{ij} \bar{\psi}_j \gamma_a \eta_L \\[5pt]
    \delta\Lambda_L&=-\frac{1}{4}\bigg(E_i\epsilon^i+E\epsilon_L \bigg)+\frac{1}{4}\gamma\cdot \bigg(T_i\epsilon^i+\mathring{T}^-\epsilon_L \bigg)\\[5pt]
    \delta E_i&=-4\bar{\epsilon}_i\cancel{D}\Lambda_L-\frac{1}{2}\varepsilon_{ij}\bigg(\bar{\epsilon}^j \mathring{\zeta}^L-\bar{\epsilon}_L\zeta^j \bigg)+\frac{1}{2}\bar{\epsilon}^j\chi_{ij}+\frac{1}{2}\bar{\epsilon}_L \mathring{\chi}_i-\frac{1}{2}\varepsilon_{ij}\bigg(\bar{E}\bar{\epsilon}^j\Lambda_L-E^j\bar{\epsilon}_L\Lambda_L \bigg) \nonumber\\
    &-4\bar{\Lambda}_L\Lambda_L\bar{\epsilon}_i\Lambda_R-4\bar{\eta}_i\Lambda_L+E u(\epsilon)_i\\
    \delta E&=-4\bar{\epsilon}_R\cancel{D}\Lambda_L-\frac{1}{2}\varepsilon_{jk}\bar{\epsilon}^j\zeta^k+\frac{1}{2}\bar{\epsilon}^j \mathring{\chi}_j+\frac{1}{2}\bar{\epsilon}_L \mathring{\chi}_L-\frac{1}{2}\varepsilon_{jk}E^k\bar{\epsilon}^j\Lambda_L-4\bar{\Lambda}_L\Lambda_L\bar{\epsilon}_R\Lambda_R\nonumber\\
    &-E_ju(\epsilon)^j-4\bar{\eta}_L\Lambda_L \\[5pt]
    \delta T_{ab}^i &=  - \bar{\epsilon}^i \cancel{D}\gamma_{ab} \Lambda_R - 4 \varepsilon^{ij} \bar{\epsilon}_j R_{ab}(Q)_R + 4 \varepsilon^{ij} \bar{\epsilon}_R R_{ab}(Q)_j + \frac{1}{8} \bar{\epsilon}_j \gamma_{ab} \chi^{ij} + \frac{1}{8} \bar{\epsilon}_R \gamma_{ab} \mathring{\chi}^i \nonumber \\
    & + \frac{1}{24} \varepsilon^{ij} \bar{\epsilon}_j \gamma _{ab} \mathring{\zeta}_R- \frac{1}{24} \varepsilon^{ij} \bar{\epsilon}_R \gamma_{ab} \zeta_j -\frac{1}{8} \varepsilon^{ij} E_j \bar{\epsilon}_R \gamma_{ab} \Lambda_R + \frac{1}{8} \varepsilon^{ij} E \bar{\epsilon}_j \gamma_{ab} \Lambda_R+ \bar{\eta}^i \gamma_{ab} \Lambda_R\nonumber\\
    & + u(\epsilon)^i \mathring{T}_{ab}^+  \\[5pt]
    \delta D^i_j&=-3\bar{\epsilon}^i \cancel{D} \zeta_j - 3 \bar{\epsilon}^i\slashed{Y}_j \mathring{\zeta}_R - 3 \varepsilon_{jk} \bar{\epsilon}^k \cancel{D} \mathring{\chi}^i + 3 \varepsilon_{jk} \bar{\epsilon}^k \slashed{Y}^i \mathring{\chi}_R - 3 \varepsilon_{jk} \bar{\epsilon}^k \slashed{Y}_l \chi^{il} + 3 \varepsilon_{jk} \bar{\epsilon}_L \cancel{D} \chi^{ik} \nonumber \\
    &  - 3 \varepsilon_{jk} \bar{\epsilon}_L \slashed{Y}^i \mathring{\chi}^k- 3 \epsilon_{jk} \slashed{Y}^k \mathring{\chi}^i + \frac{1}{4} \epsilon_{jk} \bar{\epsilon}^i \zeta^k \bar{E} - \frac{1}{4} \varepsilon_{jk} \bar{\epsilon}^i \mathring{\zeta}_L E^k+ \frac{1}{2} \varepsilon_{jk} \bar{\epsilon}^k \mathring{\zeta}_L E^i - \frac{1}{2} \varepsilon_{jk} \bar{\epsilon}_L \zeta^k E^i \nonumber \\  
    &+ \frac{3}{4} \bar{\epsilon}^i \chi_{jk}E^k + \frac{3}{4} \bar{\epsilon}^i \mathring{\chi}_j \bar{E} + 3 \bar{\epsilon}^i \gamma \cdot T_j \overset{\leftrightarrow}{\cancel{D}}\Lambda_R + 3 \bar{\epsilon}^i \gamma \cdot  \mathring{T}^- \slashed{Y}_j \Lambda_R - \bar{\epsilon}^i \cancel{D}\Lambda_R E_j - 3 \bar{\epsilon}^i \cancel{D}E_j \Lambda_R \nonumber \\
    &+ 3 \bar{\epsilon}^i \slashed{Y}_j E \Lambda_R  +\frac{3}{4} \varepsilon_{jk} \bar{E} \bar{\epsilon}^k \Lambda_L E^i - \frac{3}{4} \varepsilon_{jk}E^k \bar{\epsilon}_L \Lambda_L E^i + 3 \varepsilon_{jk} T^i \cdot \mathring{T}^+ \bar{\epsilon}^k \Lambda_L \nonumber \\ 
    & - 3 \varepsilon_{jk} T^i \cdot T^k \bar{\epsilon}_L \Lambda_L - 2 \bar{\epsilon}^i \Lambda_L \Lambda_R \zeta_j  - 3 \bar{\epsilon}^i \Lambda_L \bar{\Lambda}_R \Lambda_R E_j +3 \bar{\epsilon}^i \gamma \cdot T_j \Lambda_L \bar{\Lambda}_R \Lambda_R + u(\epsilon)^i \mathring{D}_j \nonumber\\
    &+\text{h.c.} -\text{trace}\\[5pt]
    \delta \chi_{ij} &=  2 \cancel{D}E_{(i} \epsilon_{j)} -2 \slashed{Y}_{(i} E \epsilon_{j)} - 8 \varepsilon_{k(i} \gamma \cdot R(V)^{k}{}_{j)} \epsilon_L + 8 \varepsilon_{k(i} \gamma^{ab} Y_a^k Y_{bj)} \epsilon_L\nonumber \\ & - 16 \varepsilon_{k(i} \gamma ^{ab} D_a Y_{b j)} \epsilon^k + \frac{4}{ \xi} \varepsilon_{k(i} \gamma ^{ab} \bar{\theta}_R R(Q)_{ab j)} \epsilon^k - \frac{4}{\xi} \varepsilon_{k(i} \varepsilon_{j)l} \gamma^{ab} \bigg( \bar{\psi}_L R(Q)_{ab}^l  \nonumber \\ 
    &- \bar{\psi}^l R(Q)_{ab L} \bigg) \epsilon^k - 2 \gamma \cdot \cancel{D} T_{(i} \epsilon _{j)} + 2 \gamma^{ab} \slashed{Y}_{(i} \mathring{T}^-_{ab} \epsilon_{j)} + \frac{1}{3} \varepsilon_{l(i} D^l_{j)} \epsilon_L-\frac{1}{3} \varepsilon_{l(i} \mathring{D}_{j)} \epsilon^l \nonumber\\
    & +\frac{1}{4} \varepsilon_{l(i} \bar{E} \gamma \cdot T_{j)} \epsilon^l - \frac{1}{4} \varepsilon_{l(i} E^l \gamma \cdot T_{j)} \epsilon_L - \frac{1}{3} \bar{\Lambda}_L \gamma_a \epsilon_{(i} \gamma ^a \zeta_{j)} + \frac{1}{4} \varepsilon_{m(i} E_{j)} E^m \epsilon_L \nonumber \\
    &- \frac{1}{4} \varepsilon_{m(i} E_{j)} \bar{E} \epsilon^m - \bar{\Lambda}_L \gamma^a \Lambda_R \gamma_a E_{(i} \epsilon_{j)} - \bar{\Lambda}_L \gamma \cdot T_{(i} \gamma^a \Lambda_R \gamma_a \epsilon_{j)}+ 2u(\epsilon)_{(i} \mathring{\chi}_{j)}\nonumber\\
    &+ 2 \gamma \cdot T_{(i} \eta_{j)} + 2 E_{(i} \eta_{j)} \\[5pt]
     \delta \zeta^i & = - 3 \varepsilon^{ij} \cancel{D} E_j \epsilon_R + 3 \varepsilon^{ij} \slashed{Y}_j E \epsilon_R +3 \varepsilon^{ij} \cancel{D}E
\epsilon_j +3 \varepsilon^{ij} \slashed{Y}^k E_k \epsilon_j + \varepsilon^{ij} \gamma \cdot \cancel{D} \mathring{T}^- \epsilon_j \nonumber \\ & +\varepsilon^{ij} \gamma^{ab} \cdot \slashed{Y}^k T_{abk} \epsilon_j - \varepsilon^{ij} \gamma \cdot \cancel{D}T_j \epsilon_R+ \varepsilon^{ij} \gamma^{ab} \cdot \slashed{Y}_j \mathring{T}^-_{ab} \epsilon_R -4 \gamma \cdot R(V)^i{}_j \epsilon^j + 2i \gamma \cdot R(v) \epsilon^i \nonumber \\
& + 4 \gamma^{ab} Y_a^i Y_{bj} \epsilon^j - 16 i \gamma \cdot R(A) \epsilon^i - \frac{1}{2} D^i_j \epsilon^j + \frac{1}{4} \mathring{D} \epsilon^i -\frac{1}{2} \mathring{D}^i \epsilon_L -\frac{3}{8} E^i \gamma \cdot T_j \epsilon^j \nonumber \\ 
& -\frac{3}{8} E^i \gamma \cdot \mathring{T}^- \epsilon_L + \frac{3}{8} \bar{E} \gamma \cdot \mathring{T}^- \epsilon^i + \frac{3}{8} E^j \gamma \cdot T_j \epsilon^i+ \frac{3}{8} E^i E \epsilon_L + \frac{3}{8} E^i E_j \epsilon^j + \frac{1}{8} E^j E_j \epsilon^i \nonumber \\ & + \frac{1}{8} \lvert E \rvert ^2 \epsilon^i - 4 \bar{\Lambda}_L \cancel{D}\Lambda_R \epsilon^i - 4 \bar{\Lambda}_R \cancel{D} \Lambda_L \epsilon^i - 3 \bar{\Lambda}_R \cancel{D} \gamma_{ab} \Lambda_L \gamma^{ab} \epsilon ^i \nonumber \\
&- 3 \bar{\Lambda}_L \gamma _{ab} \cancel{D} \Lambda_R \gamma ^{ab} \epsilon^i + \frac{1}{2} \varepsilon^{ij} \bar{\Lambda}_L \gamma^a \epsilon_j \gamma_a \mathring{\zeta}_R - \frac{1}{2} \varepsilon^{ij} \bar{\Lambda}_L \gamma^a \epsilon_R \gamma_a \zeta_j - 6 \bar{\Lambda}_L \Lambda_L \bar{\Lambda}_R \Lambda_R \epsilon^i \nonumber \\
& + u(\epsilon)^i \mathring{\zeta}_L+ \varepsilon^{ij} \gamma \cdot T_j \eta_L - \varepsilon^{ij} \gamma \cdot \mathring{T}^- \eta_j - 3 \varepsilon^{ij} E_j \eta_L + 3 \varepsilon^{ij} E \eta_j  \\
\delta \xi & = -\bar{\epsilon}_R \theta_R + \varepsilon_{jk} \bar{\epsilon}^j \psi^k \\
\delta \psi_i &= -\frac{1}{2} \gamma \cdot \mathcal{F}^+ \epsilon_i - 2 \varepsilon_{ij} \cancel{D} \bar{\xi} \epsilon^j - 2 \varepsilon_{ij} \slashed{Y}^j \bar{\xi} \epsilon_L  -\frac{1}{4} E_i \bar{\xi} \epsilon_R + \frac{1}{2} \bar{\Lambda}_L \theta_L \epsilon_i \nonumber \\
&+ \frac{1}{2} \varepsilon_{ij} \gamma_a \epsilon^j \bar{\Lambda}_R \gamma^a \Lambda_L \bar{\xi} + u(\epsilon)_i \psi_R - 2 \varepsilon_{ij} \bar{\xi} \eta^j  \\
\delta \psi_R& = -\frac{1}{2} \gamma \cdot \mathcal{F}^+ \epsilon_R +2 \epsilon_{jk} \slashed{Y}^k \bar{\xi} \epsilon^j - \frac{1}{4} E \bar{\xi} \epsilon_R+ \frac{1}{2} \bar{\Lambda}_L \theta_L \epsilon_R  - u(\epsilon)^i \psi_i \\
\delta \theta_L & = 2 \slashed{Y}^i \bar{\xi} \epsilon_i - 2 \cancel{D}\bar{\xi} \epsilon_R - \gamma^a \bar{\Lambda}_L \gamma_a \Lambda_R \bar{\xi} \epsilon_R - \bar{\Lambda}_R \psi_i \epsilon^i - \frac{1}{4} \varepsilon_{ij} E^i \bar{\xi} \epsilon^j -  \bar{\Lambda}_R \psi_R \epsilon_L   - 2 \bar{\xi} \eta_L 
\end{align}
\end{subequations}}
\section{$SU(2) \times U(1)$ Dilaton Weyl multiplet}\label{dil2}

Recently, in \cite{Ciceri:2024xxf}, a dilaton Weyl multiplet was constructed for \(\NN = 4\) conformal supergravity in four dimensions which had the R-symmetry group $USp(4)$. Preliminary investigations suggest that the supersymmetric truncation of this Weyl multiplet to $\NN=3$ will give us an $\NN=3$ dilaton Weyl multiplet with R-symmetry $SU(2)\times U(1)\subset USp(4)$. However, the $\NN=3$ dilaton Weyl multiplet constructed in section-\ref{1st dilaton} has R-symmetry $SU(2)\times U(1)\times U(1)$. Hence, in order to connect with the $\NN=4$ dilaton Weyl multiplet constructed in \cite{Ciceri:2024xxf}, we need to explicitly break one of the $U(1)$ using an already existing field of the dilaton Weyl multiplet and thereby obtaining an $\NN=3$ dilaton Weyl multiplet with R-symmetry $SU(2)\times U(1)$. We use the field $\xi$ for this purpose and impose the gauge fixing condition
\begin{align}\label{real}
     \xi&= \bar{\xi}
\end{align}
Demanding $\delta (\xi-\bar{\xi})=0$ (where $\delta$ is the set of all gauge transformations including supersymmetry), will give us the following relations between the transformation parameters: 
\begin{align}
\lambda_A+\lambda_v=\frac{1}{2i\xi}\left(-\bar{\epsilon}_R\theta_R+\varepsilon_{jk}\bar{\epsilon}^{j}\psi^{k}-\text{h.c}\right)
\end{align}
We can use the above equation to write the transformation parameter $\lambda_v$ in terms of the $Q$-supersymmetry parameters and the $U(1)_A$ parameter $\lambda_A$ and  interpret it as a modification in $Q$-supersymmetry as well as $U(1)_A$ as shown below:
\begin{align}\label{mod_Qsusy_2}\delta_Q^{\text{new}}(\epsilon)&=\delta_Q(\epsilon)+\delta_{U(1)_v}(v(\epsilon))\;,\nonumber \\
    \delta_{U(1)_A}^{\text{new}}(\lambda_A)&=\delta_{U(1)_A}(\lambda_A)+\delta_{U(1)_v}(-\lambda_A)\;,
    \end{align}
where,
\begin{align}
    v(\epsilon)=\frac{1}{2i\xi}\left(-\bar{\epsilon}_R\theta_R+\varepsilon_{jk}\bar{\epsilon}^{j}\psi^{k}-\text{h.c}\right)\;,
\end{align}
 and $\delta_Q$ appearing in the R.H.S of \eqref{mod_Qsusy_2} is the $\delta_Q$ appearing in section \ref{1dilaton4d}. As a result of which the weights of the fields under this redefined $U(1)_A$ will become $c_v-c_A$ where $c_v$ and $c_A$ are the weights, under $U(1)_v$ and $U(1)_A$ respectively, of the fields of the dilaton Weyl multiplet derived section \ref{1st dilaton} and as tabulated in Table-\eqref{1dilaton4d}.

The modified gauge transformations of the $U(1)_v$ dependent gauge field $v_a$ will acquire terms depending on the derivative of the $Q$-supersymmetry parameters as well as the $U(1)_A$ parameter. Hence a redefined field, as shown below, will become covariant:
\begin{align}
\mathring{K}_a=\mathring{v}_a+A_a-\frac{1}{2}v(\psi_a)
\end{align}
Using \eqref{vasol} as well as the gauge fixing conditions \eqref{real}, this can be re-written in a manifestly covariant form as shown below:
\begin{align}
\mathring{K}_a&= \frac{i}{2\xi^2} \bigg(- \frac{1}{3!} \varepsilon_{abcd} H^{bcd} +\frac{1}{2}\bar{\theta}_R\gamma_a\theta_L+\frac{1}{2}\bar{\psi}^j\gamma_a\psi_j+\frac{1}{2}\bar{\psi}_L\gamma_a\psi_R + \frac{1}{2} \xi^2 \bar{\Lambda}_R \gamma_a \Lambda_L \bigg)
\end{align}
The covariant derivatives appearing in the dilaton Weyl multiplet constructed in section-\ref{1st dilaton} will also split up into a new covariant derivative which will be covariant with respect to all the modified gauge transformations and terms depending on the dependent covariant field $\mathring{K}_a$. For example, let us consider a field $\Psi$ which transforms with a weight $c_v$ under $U(1)_v$ in the dilaton Weyl multiplet derived in section \ref{1st dilaton}. The covariant derivative will split up as:
\begin{align}
    D_{\mu}\Psi=D_{\mu}^{\text{new}}\Psi-ic_v\mathring{K}_{\mu}\Psi
\end{align}
Henceforth, in this section we will be using all the new objects (transformations as well as covariant derivatives) and hence drop the superscript ``new'' for the sake of simplicity. In this way, we arrive at a dilaton Weyl multiplet with R-symmetry $SU(2)\times U(1)$, where the fields of the multiplet along with all their properties are tabulated in Table-\eqref{2dilaton4d}. 
\begin{table}[h!]
		\centering
		\centering
		\begin{tabular}{|c|c|c|c|c|}
			\hline
			Field&Properties&SU(2) irreps&$w$ & $c_A$\\
			\hline
			\multicolumn{5}{|c|}{Independent Gauge fields}\\
			\hline
			$e_\mu^a$&vielbein&\bf{1}&$-1$&0\\
			$\psi_\mu^i$, & $\gamma_5 \psi_\mu^i=\psi_\mu^i$,  &\bf{2}&$-1/2$&$0$\\
            $\psi_{\mu L}$, & $\gamma_5 \psi_\mu=\psi_{\mu L}$ &\bf{1}&$-1/2$&$-3/2$\\
			$v_\mu{}^{i}{}_j $&$(v_\mu{}^i{}_j)^*\equiv v_{\mu i}{}^j=-v_\mu{}^j{}_i $&\bf{3}&0&0\\
   			$A_\mu$& $U(1)_A$ gauge field&\bf{1}&0&0\\
                  $b_\mu$&Dilatation gauge field&\bf{1}&$0$&$0$\\
      $B_{\mu \nu}$& Two-form gauge field& \bf{1}&0&0\\
            $C_\mu$ & $ U(1) $ gauge field&\bf{1}&0&0\\
              $\Tilde{C}_\mu$ & $ U(1) $ gauge field&\bf{1}&0&0\\
			\hline
			\multicolumn{5}{|c|}{Covariant fields}\\
			\hline
   $Y_a^i$&Boson&\bf{2}&1&$1/2$\\
 $  T_{ab}^i$& $T^i_{ab}=\frac{1}{2}\varepsilon_{abcd}T^{icd} $&\bf{1}&1&$3/2$\\
 $E_i$& Complex & $\bf{{2}}$ &1&$-3/2$\\
  $E$& Complex & $\bf{1}$ &1&0\\
$D^i{}_j$& $(D^i{}_j)^*\equiv D_i{}^j=D^j{}_i $, $D^i{}_i=0$& \bf{3}&2&0\\ 
$\chi_{ij} $&$\gamma_5\chi_{ij} =\chi_{ij} $& $\bf{{3}}$&$3/2$&$-3/2$\\
$\Lambda_L$&$\gamma_5\Lambda_L=\Lambda_L$&\bf{1}&$1/2$&$-3/2$\\
$\zeta^i$& $\gamma_5\zeta^i=\zeta^i$&\bf{2}& $3/2$&0\\
$\xi $& Real scalar & \bf{1} & $1$& $0$\\
$\psi_i$&$\gamma_5 \psi_i = - \gamma_5\psi_i$ &\bf{2}&$3/2$&0\\
$\psi_R$& $\gamma_5 \psi_R = - \psi_R$& \bf{1}&$3/2$&$3/2$\\
$\theta_L $& $\gamma_5 \theta_L = \theta_L$ & \bf{1}&$3/2$&$3/2$\\
 			\hline
		\end{tabular}
		\caption{Field content of the $\mathcal{N}=3$ dilaton Weyl multiplet with R-symmetry $SU(2)\times U(1)$}
		\label{2dilaton4d}	
	\end{table}
The $Q$ and $S$-supersymmetry transformations are given in \eqref{2dilatontrans}.
\allowdisplaybreaks
{\begin{subequations}\label{2dilatontrans}
\begin{align}
\delta e_\mu^a&=\bar{\epsilon}_i\gamma^a\psi_\mu^i+\bar{\epsilon}_R\gamma^a\psi_{\mu L}+\text{h.c.} \\[5pt]
\delta\psi_\mu^i&=-\frac{i}{2}v(\epsilon)\psi^i_\mu+2\mathcal{D}_\mu\epsilon^i +i \mathring{K}_\mu \epsilon^i +  \frac{i}{2}v(\psi) \epsilon^i -2Y^i_\mu\epsilon_L-u(\psi_\mu)^i\epsilon_L-\frac{1}{8} \varepsilon^{ij} \gamma\cdot (T_j\gamma_\mu\epsilon_R \nonumber\\
    &- \mathring{T}^-\gamma_\mu\epsilon_j )-\varepsilon^{ij}\bar{\epsilon}_j\psi_{\mu R}\Lambda_L +\varepsilon^{ij} \bar{\epsilon}_R \psi_{\mu j}\Lambda_L+u(\epsilon)^i\psi_{\mu L}-\gamma_\mu \eta^i \\[5pt]
\delta \psi_{\mu L}&= i v(\epsilon) \psi_{\mu L}+2\mathcal{D}_\mu\epsilon_L - 2i \mathring{K}_\mu \epsilon_L -\frac{i}{2}v(\psi) \epsilon_L  +2Y_{\mu j}\epsilon^j + u(\psi_\mu)_i \epsilon^i -\frac{1}{8}\varepsilon^{ij}\gamma\cdot T_i\gamma_\mu \epsilon_j \nonumber \\
&-\varepsilon^{ij}\bar{\epsilon}_i\psi_{\mu j}\Lambda_L-u(\epsilon)_i\psi^i_\mu  -\gamma_\mu\eta_R\\[5pt]
     \delta v_\mu{}^i{}_j& =\bar{\epsilon}^i\phi_{\mu j}-\frac{1}{48} \bar{\epsilon}^i \gamma_\mu \zeta_j +\frac{1}{16}\bigg(\varepsilon_{jk}\bar{\epsilon}^k\gamma_\mu \mathring{\chi}^i-\varepsilon_{jk}\bar{\epsilon}_L\gamma_\mu\chi^{ik} \bigg)+\frac{1}{8}E^i\varepsilon_{jk}\bigg(\bar{\epsilon}^k\psi_{\mu L}-\bar{\epsilon}_L\psi^k_\mu \bigg)\nonumber\\
    &-\frac{1}{16}\bar{\epsilon}^i\gamma\cdot T_j\gamma_\mu\Lambda_R-\frac{1}{16}\bar{\epsilon}^i\gamma_\mu\Lambda_RE_j+\frac{1}{4}\bar{\epsilon}^i\gamma^a\psi_{\mu j}\bar{\Lambda}_L\gamma_a\Lambda_R-u(\epsilon)^i\bigg(Y_{\mu j}+\frac{1}{2} u(\psi_\mu)_j \bigg)\nonumber\\
    &-\bar{\psi}_\mu^i\eta_j -\text{h.c.}-\text{trace} \\
    \delta b_\mu&=\frac{1}{2}\bigg(\bar{\epsilon}^i\phi_{\mu i}+\bar{\epsilon}_L\phi_{\mu L}-\bar{\psi}_\mu^i\eta_i-\bar{\psi}_{\mu L}\eta_L \bigg)+\text{h.c.} \\[5pt]
    \delta A_\mu&= \frac{i}{6}\bigg( \bar{\epsilon}^i\phi_{\mu i}+\bar{\epsilon}_L\phi_{\mu L}\bigg)+\frac{i}{36}\bigg( \bar{\epsilon}^i\gamma_\mu\zeta_i+\bar{\epsilon}_L\gamma_\mu \mathring{\zeta}_R\bigg)+\frac{i}{12}\varepsilon_{kl}\bigg(\bar{E}\bar{\epsilon}^k\psi^l_\mu-E^l\bar{\epsilon}^k\psi_{\mu L}\nonumber\\
    &+E^l\bar{\epsilon}_L\psi^k_\mu
    \bigg) +\frac{i}{12}\bigg(\bar{\epsilon}^i\gamma\cdot T_i\gamma_\mu\Lambda_R+\bar{\epsilon}_L\gamma\cdot \mathring{T}^-\gamma_\mu \Lambda_R \bigg)+\frac{i}{12}\bigg(E_i\bar{\epsilon}^i\gamma_\mu\Lambda_R+E\bar{\epsilon}_L\gamma_\mu \Lambda_R \bigg) \nonumber\\
    &-\frac{i}{3}\bigg( \bar{\epsilon}^i\gamma^a\psi_{\mu i}+\bar{\epsilon}_L\gamma_a\psi_{\mu R}\bigg)\bar{\Lambda}_L\gamma_a\Lambda_R-\frac{i}{16}\bigg(\bar{\psi}_\mu^i\eta_i+\psi_{\mu L}\eta_L \bigg)+\text{h.c.}\\[5pt]
    \delta{C}_{\mu}&= \bar{\epsilon}^{i}\gamma_{\mu}\psi_{i}+ \bar{\epsilon}_{L}\gamma_{\mu}\psi_{R} -2\bar{\epsilon}^{i}\psi_{\mu}^j \xi \varepsilon_{ij} - \bar{\epsilon}_{R}\gamma_{\mu}\Lambda_L {{\xi}} +\text{h.c.}\\[5pt]
      \delta\Tilde{C}_\mu&= -i\bar{\epsilon}^i\gamma_\mu\psi_i-i\bar{\epsilon}_L\gamma_\mu \psi_R+2i\bar{\epsilon}^i\psi_\mu^j\varepsilon_{ij}{\xi}+i\bar{\epsilon}_R\gamma_\mu\Lambda_L \xi+\text{h.c.} \\[5pt]
      \delta B_{\mu \nu} &= (\bar{\epsilon}_R \gamma_{\mu \nu} \theta_R \xi + \varepsilon_{ij} \bar{\epsilon}^i \gamma_{\mu \nu} \psi^j \xi - 2 \xi^2 \bar{\epsilon}_R \gamma_{[\mu} \psi_{\nu] L} + 2 \xi^2 \bar{\epsilon}^i \gamma_{[\mu} \psi_{\nu ]i} + \text{h.c}) \nonumber\\
      &+ \frac{1}{2} C_{[\mu} \delta C_{\nu]}  + \frac{1}{2} \Tilde{C}_{[\mu} \delta  \Tilde{C}_{\nu]} - \frac{\lambda}{4} F_{\mu \nu} - \frac{\Tilde{\lambda}}{4} G_{\mu \nu} + 2 \partial_{[\mu} \Lambda_{\nu ]}\\[5pt]
       \delta Y_a^i& = \bar{\epsilon}^i D_a(\frac{\theta_L}{{\xi}})- \varepsilon^{ij} \bar{\epsilon}_j D_a(\frac{\psi_R}{\xi})+ \varepsilon^{ij} \bar{\epsilon}_R D_a(\frac{\psi_j}{\xi}) -\frac{1}{48} \bar{\epsilon}^i \gamma_a \mathring{\zeta}_R + \frac{1}{48} \bar{\epsilon}_R \gamma_a \zeta^i + \frac{1}{16} \varepsilon_{jk} \bar{\epsilon}^j \gamma_a \chi^{ik} \nonumber \\& - \frac{1}{16} \varepsilon^{ij} \bar{\epsilon}_j \gamma_a \mathring{\chi}_L + \frac{1}{16} \varepsilon^{ij} \bar{\epsilon}_R \gamma_a \mathring{\chi}_j - \frac{1}{16} \bar{\epsilon}^i \gamma \cdot \mathring{T}^- \gamma_a \Lambda_R + \frac{1}{16} \bar{\epsilon}_R \gamma \cdot T^i \gamma_a \Lambda_L - \frac{1}{16} \bar{\epsilon}^i \gamma_a \Lambda_R E \nonumber \\
    & + \frac{1}{16} \bar{\epsilon}_R \gamma_a \Lambda_L E^i - \frac{1}{2 {\xi}} \bar{\theta}_L \bigg( -2Y_a^i \epsilon_L -\frac{1}{8} \varepsilon^{ij}  \gamma \cdot T_j \gamma_a \epsilon_R + \frac{1}{8} \varepsilon^{ij}  \gamma \cdot \mathring{T}^- \gamma_a \epsilon_j \nonumber \bigg) \\
    &-\frac{1}{{\xi}} \varepsilon^{ij}Y_{aj}\bar{\psi}_R \epsilon_R  + \frac{1}{2{\xi}} \bar{\psi}_R  \bigg( \frac{1}{8} \gamma \cdot T^i \gamma_a \epsilon_L - \frac{1}{8} \gamma \cdot \mathring{T}^+ \gamma_a \epsilon^i \bigg) + \frac{1}{8 {\xi}} \bar{\psi}_j \gamma \cdot T^{[i} \gamma_a \epsilon^{j]} \nonumber \\
    &-\frac{1}{{\xi}} \varepsilon^{ij} \bar{\psi}_j \epsilon_k Y_a^k +  i \bar{\epsilon}^i \mathring{K}_a  \frac{\theta_L}{\xi} - 2i \varepsilon^{ij} \bar{\epsilon}_j \mathring{K}_a \frac{\psi_R}{\xi} +\frac{i}{2} \varepsilon^{ij} \bar{\epsilon}_R \mathring{K}_a \frac{\psi_j}{ \xi}   + \frac{1}{2 \xi} \bar{\theta}_L \gamma_a \eta^i - \frac{1}{2\xi} \varepsilon^{ij} \bar{\psi}_R \gamma_a \eta_j\nonumber\\
    &+ \frac{1}{2\xi} \epsilon^{ij} \bar{\psi}_j \gamma_a \eta_L \\[5pt]
    \delta\Lambda_L&=-\frac{1}{4}\bigg(E_i\epsilon^i+E\epsilon_L \bigg)+\frac{1}{4}\gamma\cdot \bigg(T_i\epsilon^i+\mathring{T}^-\epsilon_L \bigg)\\[5pt]
    \delta E_i&= \frac{i}{2} v(\epsilon) E_i -4\bar{\epsilon}_i\cancel{D}\Lambda_L-\frac{1}{2}\varepsilon_{ij}\bigg(\bar{\epsilon}^j \mathring{\zeta}^L-\bar{\epsilon}_L\zeta^j \bigg)+\frac{1}{2}\bar{\epsilon}^j\chi_{ij}+\frac{1}{2}\bar{\epsilon}_L \mathring{\chi}_i-\frac{1}{2}\varepsilon_{ij}\bigg(\bar{E}\bar{\epsilon}^j\Lambda_L\nonumber\\
    &-E^j\bar{\epsilon}_L\Lambda_L \bigg) -4\bar{\Lambda}_L\Lambda_L\bar{\epsilon}_i\Lambda_R-4\bar{\eta}_i\Lambda_L+E u(\epsilon)_i\\[5pt]
    \delta E&= -i v(\epsilon) E -4\bar{\epsilon}_R\cancel{D}\Lambda_L-\frac{1}{2}\varepsilon_{jk}\bar{\epsilon}^j\zeta^k+\frac{1}{2}\bar{\epsilon}^j \mathring{\chi}_j+\frac{1}{2} \bar{\epsilon}_L \mathring{\chi}_L -\frac{1}{2}\varepsilon_{jk}E^k\bar{\epsilon}^j\Lambda_L-4\bar{\Lambda}_L\Lambda_L\bar{\epsilon}_R\Lambda_R \nonumber\\
    &-E_ju(\epsilon)^j-4\bar{\eta}_L\Lambda_L\\[5pt]
    \delta T_{ab}^i &= -\frac{i}{2} v(\epsilon) T_{ab}^i  - \bar{\epsilon}^i \cancel{D}\gamma_{ab} \Lambda_R - 4 \varepsilon^{ij} \bar{\epsilon}_j R_{ab}(Q)_R + 4 \varepsilon^{ij} \bar{\epsilon}_R R_{ab}(Q)_j + \frac{1}{8} \bar{\epsilon}_j \gamma_{ab} \chi^{ij} + \frac{1}{8} \bar{\epsilon}_R \gamma_{ab} \mathring{\chi}^i  \nonumber \\
    &+ \frac{1}{24} \varepsilon^{ij} \bar{\epsilon}_j \gamma _{ab} \mathring{\zeta}_R - \frac{1}{24} \varepsilon^{ij} \bar{\epsilon}_R \gamma_{ab} \zeta_j -\frac{1}{8} \varepsilon^{ij} E_j \bar{\epsilon}_R \gamma_{ab} \Lambda_R + \frac{1}{8} \varepsilon^{ij} E \bar{\epsilon}_j \gamma_{ab} \Lambda_R+ \bar{\eta}^i \gamma_{ab} \Lambda_R \nonumber\\
    &+ u(\epsilon)^i \mathring{T}_{ab}^+  \\[5pt]
    \delta D^i_j&= -3\bar{\epsilon}^i \cancel{D} \zeta_j - 3 \bar{\epsilon}^i\slashed{Y}_j \mathring{\zeta}_R - 3 \varepsilon_{jk} \bar{\epsilon}^k \cancel{D} \mathring{\chi}^i + 3 \varepsilon_{jk} \bar{\epsilon}^k \slashed{Y}^i \mathring{\chi}_R - 3 \varepsilon_{jk} \bar{\epsilon}^k \slashed{Y}_l \chi^{il} + 3 \varepsilon_{jk} \bar{\epsilon}_L \cancel{D} \chi^{ik}  \nonumber \\
    & - 3 \varepsilon_{jk} \bar{\epsilon}_L \slashed{Y}^i \mathring{\chi}^k- 3 \epsilon_{jk} \slashed{Y}^k \mathring{\chi}^i + \frac{1}{4} \epsilon_{jk} \bar{\epsilon}^i \zeta^k \bar{E} - \frac{1}{4} \varepsilon_{jk} \bar{\epsilon}^i \mathring{\zeta}_L E^k+ \frac{1}{2} \varepsilon_{jk} \bar{\epsilon}^k \mathring{\zeta}_L E^i - \frac{1}{2} \varepsilon_{jk} \bar{\epsilon}_L \zeta^k E^i \nonumber \\ 
    &+ \frac{3}{4} \bar{\epsilon}^i \chi_{jk}E^k + \frac{3}{4} \bar{\epsilon}^i \mathring{\chi}_j \bar{E} + 3 \bar{\epsilon}^i \gamma \cdot T_j \overset{\leftrightarrow}{\cancel{D}}\Lambda_R + 3 \bar{\epsilon}^i \gamma \cdot  \mathring{T}^- \slashed{Y}_j \Lambda_R - \bar{\epsilon}^i \cancel{D}\Lambda_R E_j - 3 \bar{\epsilon}^i \cancel{D}E_j \Lambda_R \nonumber \\
    &+ 3 \bar{\epsilon}^i \slashed{Y}_j E \Lambda_R  +\frac{3}{4} \varepsilon_{jk} \bar{E} \bar{\epsilon}^k \Lambda_L E^i - \frac{3}{4} \varepsilon_{jk}E^k \bar{\epsilon}_L \Lambda_L E^i + 3 \varepsilon_{jk} T^i \cdot \mathring{T}^+ \bar{\epsilon}^k \Lambda_L  \nonumber \\ 
    &- 3 \varepsilon_{jk} T^i \cdot T^k \bar{\epsilon}_L \Lambda_L  - 2 \bar{\epsilon}^i \Lambda_L \Lambda_R \zeta_j - 3 \bar{\epsilon}^i \Lambda_L \bar{\Lambda}_R \Lambda_R E_j +3 \bar{\epsilon}^i \gamma \cdot T_j \Lambda_L \bar{\Lambda}_R \Lambda_R + u(\epsilon)^i \mathring{D}_j \nonumber\\
    & +\frac{3i}{2}\bar
    {\epsilon}^i\slashed{\mathring{K}}\zeta_j+\frac{3i}{2} \varepsilon_{jk} \bar{\epsilon}^k \slashed{\mathring{K}} \mathring{\chi}^i +3\varepsilon_{jk}\bar{\epsilon} \slashed{\mathring{K}}\chi^{ik} +\frac{3i}{2}\bar{\epsilon}^i \gamma^{ab} \slashed{\mathring{K}}T_{abj}+\frac{3i}{2} \epsilon^i \slashed{\mathring{K}} E_j \Lambda_R  +\text{h.c.} -\text{trace} \\[5pt]
    \delta \chi_{ij} &= i v(\epsilon) \chi_{ij} + 2 \cancel{D}E_{(i} \epsilon_{j)} -2 \slashed{Y}_{(i} E \epsilon_{j)} - 8 \varepsilon_{k(i} \gamma \cdot R(V)^{k}{}_{j)} \epsilon_L+ 8 \varepsilon_{k(i} \gamma^{ab} Y_a^k Y_{bj)} \epsilon_L \nonumber \\
    & - 16 \varepsilon_{k(i} \gamma ^{ab} D_a Y_{b j)} \epsilon^k + \frac{4}{ \xi} \varepsilon_{k(i} \gamma ^{ab} \bar{\theta}_R R(Q)_{ab j)} \epsilon^k - \frac{4}{\xi} \varepsilon_{k(i} \varepsilon_{j)l} \gamma^{ab} \bigg( \bar{\psi}_L R(Q)_{ab}^l  \nonumber \\ 
    &- \bar{\psi}^l R(Q)_{ab L} \bigg) \epsilon^k - 2 \gamma \cdot \cancel{D} T_{(i} \epsilon _{j)} + 2 \gamma^{ab} \slashed{Y}_{(i} \mathring{T}^-_{ab} \epsilon_{j)} + \frac{1}{3} \varepsilon_{l(i} D^l_{j)} \epsilon_L-\frac{1}{3} \varepsilon_{l(i} \mathring{D}_{j)} \epsilon^l \nonumber\\
    & +\frac{1}{4} \varepsilon_{l(i} \bar{E} \gamma \cdot T_{j)} \epsilon^l - \frac{1}{4} \varepsilon_{l(i} E^l \gamma \cdot T_{j)} \epsilon_L - \frac{1}{3} \bar{\Lambda}_L \gamma_a \epsilon_{(i} \gamma ^a \zeta_{j)} + \frac{1}{4}\left( \varepsilon_{m(i} E_{j)} E^m \epsilon_L\right. \nonumber \\
    &\left.- \varepsilon_{m(i} E_{j)} \bar{E} \epsilon^m\right)   - \bar{\Lambda}_L \gamma^a \Lambda_R \gamma_a E_{(i} \epsilon_{j)} - \bar{\Lambda}_L \gamma \cdot T_{(i} \gamma^a \Lambda_R \gamma_a \epsilon_{j)}+ 2u(\epsilon)_{(i} \mathring{\chi}_{j)} - i \slashed{\mathring{K}} E_{(i} \epsilon_{j)} 
 \nonumber\\
    &+ i \gamma \cdot \slashed{\mathring{K}}T_{(i} \epsilon_{j)} + 2 \gamma \cdot T_{(i} \eta_{j)} + 2 E_{(i} \eta_{j)}  \\
     \delta \zeta^i & =-\frac{i}{2} v(\epsilon) \zeta^i  - 3 \varepsilon^{ij} \cancel{D} E_j \epsilon_R + 3 \varepsilon^{ij} \slashed{Y}_j E \epsilon_R +3 \varepsilon^{ij} \cancel{D}E
\epsilon_j +3 \varepsilon^{ij} \slashed{Y}^k E_k \epsilon_j + \varepsilon^{ij} \gamma \cdot \cancel{D}\mathring{T}^- \epsilon_j \nonumber \\ & + \varepsilon^{ij} \gamma \cdot \slashed{Y}^k T_k \epsilon_j - \varepsilon^{ij} \gamma \cdot \cancel{D}T_j \epsilon_R+ \varepsilon^{ij} \gamma \cdot \slashed{Y}_j \mathring{T}^- \epsilon_R-4 \gamma \cdot R(V)^i{}_j \epsilon^j + 4i \gamma^{ab} D_{[a} \mathring{K}_{b]} \epsilon^i  \nonumber \\
& + \frac{1}{4 \xi}\gamma^{ab} \bigg(- \bar{\theta}_R {R(Q)}_{ab R}+ \varepsilon_{jk} \bar{R(Q)_{ab}^j \psi^k -\text{h.c}}   \bigg) \epsilon^i- 4 \gamma^{ab} Y_a^i Y_{bj} \epsilon^j - 18 i \gamma \cdot R(A) \epsilon^i \nonumber \\ 
& - \frac{1}{2} D^i_j \epsilon^j + \frac{1}{4} \mathring{D} \epsilon^i -\frac{1}{2} \mathring{D}^i \epsilon_L -\frac{3}{8} E^i \gamma \cdot T_j \epsilon^j-\frac{3}{8} E^i \gamma \cdot \mathring{T}^- \epsilon_L + \frac{3}{8} \bar{E} \gamma \cdot \mathring{T}^- \epsilon^i + \frac{3}{8} E^j \gamma \cdot T_j \epsilon^i \nonumber \\ 
& + \frac{3}{8} E^i E \epsilon_L + \frac{3}{8} E^i E_j \epsilon^j + \frac{1}{8} E^j E_j \epsilon^i+ \frac{1}{8} \lvert E \rvert ^2 \epsilon^i - 4 \bar{\Lambda}_L \cancel{D}\Lambda_R \epsilon^i - 4 \bar{\Lambda}_R \cancel{D} \Lambda_L \epsilon^i  \nonumber \\ & - 3 \bar{\Lambda}_R \cancel{D} \gamma_{ab} \Lambda_L \gamma^{ab} \epsilon ^i - 3 \bar{\Lambda}_L \gamma _{ab} \cancel{D} \Lambda_R \gamma ^{ab} \epsilon^i + \frac{1}{2} \varepsilon^{ij} \bar{\Lambda}_L \gamma^a \epsilon_j \gamma_a \mathring{\zeta}_R - \frac{1}{2} \varepsilon^{ij} \bar{\Lambda}_L \gamma^a \epsilon_R \gamma_a \zeta_j \nonumber \\
& - 6 \bar{\Lambda}_L \Lambda_L \bar{\Lambda}_R \Lambda_R \epsilon^i + u(\epsilon)^i \mathring{\zeta}_L+ \frac{3i}{2} \varepsilon^{ij} \slashed{\mathring{K}}E_j \epsilon_R+3i \varepsilon^{ij}\slashed{\mathring{K}}E \epsilon_j +i\varepsilon^{ij} \gamma^{ab} \slashed{\mathring{K}} \mathring{T}^- _{ab} \epsilon_j \nonumber\\
&+ \frac{i}{2} \varepsilon^{ij} \slashed{\mathring{K}} T_j \epsilon_R + \varepsilon^{ij} \gamma \cdot T_j \eta_L - \varepsilon^{ij} \gamma \cdot \mathring{T}^- \eta_j - 3 \varepsilon^{ij} E_j \eta_L + 3 \varepsilon^{ij} E \eta_j  \\[5pt]
\delta \xi & = -   \bar{\epsilon}_R \theta_R + \varepsilon_{jk} \bar{\epsilon}^j \psi^k + h.c \\[5pt]
\delta \psi_i &= \frac{i}{2} v(\epsilon) \psi_i -\frac{1}{2} \gamma \cdot \mathcal{F}^+ \epsilon_i - 2 \varepsilon_{ij} \cancel{D} \xi \epsilon^j - 2 \varepsilon_{ij} \slashed{Y}^j \xi \epsilon_L  -\frac{1}{4} E_i \xi \epsilon_R + \frac{1}{2} \bar{\Lambda}_L \theta_L \epsilon_i \nonumber \\
&+2i \varepsilon_{jk} \slashed{\mathring{K}} \xi \epsilon^j + \frac{1}{2} \varepsilon_{ij} \gamma_a \epsilon^j \bar{\Lambda}_R \gamma^a \Lambda_L \xi + u(\epsilon)_i \psi_R - 2 \varepsilon_{ij} \xi \eta^j \\[5pt]
\delta \psi_R& =-i v(\epsilon) \psi_R -\frac{1}{2} \gamma \cdot \mathcal{F}^+ \epsilon_R +2 \epsilon_{jk} \slashed{Y}^k \xi \epsilon^j - \frac{1}{4} E \xi \epsilon_R+ \frac{1}{2} \bar{\Lambda}_L \theta_L \epsilon_R  - u(\epsilon)^i \psi_i \\[5pt]
\delta \theta_L & = 2 \slashed{Y}^i \xi \epsilon_i - 2 \cancel{D}\xi \epsilon_R - \gamma^a \bar{\Lambda}_L \gamma_a \Lambda_R \xi \epsilon_R - \bar{\Lambda}_R \psi_i \epsilon^i + 2i \slashed{\mathring{K}} \xi \epsilon_R - \bar{\Lambda}_R \psi_R \epsilon_L-\frac{1}{4} \varepsilon_{ij} E^i \xi \epsilon^j \nonumber\\
& - 2 \xi \eta_L 
\end{align}
    \end{subequations}}
The crucial differences between the $SU(2)\times U(1)$ dilaton Weyl multiplet derived in this section and the $SU(2)\times U(1)\times U(1)$ dilaton Weyl multiplet of section-\ref{1st dilaton} are as follows. The field $\xi$ here is real, whereas it is complex in section-\ref{1st dilaton}. The composite gauge field $\mathring{v}_{\mu}$ corresponding to one of the $U(1)$ in section-\ref{1st dilaton} is replaced by a composite covariant field $\mathring{K}_a$ in this section. Although we do not show explicitly, one can arrive at the dilaton Weyl multiplet derived in this section from a supersymmetric truncation of the $\NN=4$ dilaton Weyl multiplet derived in \cite{Ciceri:2024xxf}.

 \section{Discussion}\label{conclusions}
In this paper, we have constructed a dilaton Weyl multiplet for $\NN=3$ conformal supergravity. We have obtained it by first coupling an on-shell vector multiplet to the standard Weyl multiplet \cite{vanMuiden:2017qsh, Hegde:2018mxv} and using the field equations of the vector multiplet to algebraically solve for some components of the standard Weyl multiplet and replacing them with other fields. The new set of fields, together with the remaining fields of the standard Weyl multiplet, constitute the dilaton Weyl multiplet. The breaking of $SU(3)\times U(1)$ R-symmetry of the standard Weyl multiplet to $SU(2)\times U(1) \times U(1)$ was essential in order to do this. 

Preliminary analysis suggests that if we supersymmetrically truncate the $\NN=4$ dilaton Weyl multiplet of \cite{Ciceri:2024xxf} to $\NN=3$, the dilaton Weyl multiplet that one would obtain will contain $SU(2)\times U(1)$ R-symmetry instead of $SU(2)\times U(1) \times U(1)$. In order to arrive at such a dilaton Weyl multiplet for $\NN=3$ conformal supergravity, we use the previously constructed $\NN=3$ dilaton Weyl multiplet which had  $SU(2)\times U(1) \times U(1)$ R-symmetry, gauge fix one of the two $U(1)$ and arrive at the $SU(2)\times U(1)$ dilaton Weyl multiplet in section \ref{dil2}. To establish the above claim that this is related to the $\NN=4$ dilaton Weyl multiplet of \cite{Ciceri:2024xxf}, carrying out an explicit supersymmetric truncation of the $\NN=4$ dilaton Weyl multiplet will be necessary. We leave it for our immediate future work. 

In \cite{Adhikari:2024esl}, a ``new'' $\NN=4$ dilaton Weyl multiplet was constructed by coupling an additional on-shell vector multiplet to the ``old'' dilaton Weyl multiplet of \cite{Ciceri:2024xxf}. This ``new'' $\NN=4$ dilaton Weyl multiplet of \cite{Adhikari:2024esl} has an R-symmetry $SU(2)\times SU(2)$. Since the new dilaton Weyl multiplet comes from coupling a vector multiplet to the old dilaton Weyl multiplet, which was itself obtained from coupling a vector multiplet to a standard Weyl multiplet, one can think of the new $\NN=4$ dilaton Weyl multiplet as a package of two $\NN=4$ 
on-shell vector multiplets and an $\NN=4$ standard Weyl multiplet. Following similar construction, in one of our recent work in $\NN=3$ conformal supergravity \cite{Adhikari:2025wwb}, we construct a new dilaton Weyl multiplet, which is obtained by coupling two on-shell vector multiplets with the $\NN=3$ standard Weyl multiplet. One may expect that this $\NN=3$ dilaton Weyl multiplet may be related to the new $\NN=4$ dilaton Weyl multiplet of \cite{Adhikari:2024esl} upon supersymmetric truncation. In this new construction, some more auxiliary fields belonging to the standard Weyl multiplet become composite. However, some more components of the auxiliary fields still remain. One may wonder if we can push this construction to another variant of the dilaton Weyl multiplet which is obtained by coupling three vector multiplets to the standard Weyl multiplet. In this construction, it is possible that all the auxiliaries belonging to the standard Weyl multiplet are completely solved and one has a dilaton Weyl multiplet without the presence of any auxiliary field. The only obstruction in this construction appears to be the solution of the fermionic field equations. The number of components of the fermionic field equations coming from the three vector multiplets would be $12\times 4=48$ which outnumbers the number of components carried by the fermionic auxiliary fields $\chi_{IJ}$ and $\zeta_I$ belonging to the standard Weyl multiplet which is $9\times 4=36$. Even after solving all the fermionic auxiliaries, one would be still left with some fermionic equations and one needs to figure out a way to solve these equations. If one can figure a way out of this subtlety and the construction of this variant of the dilaton Weyl multiplet is possible, this would have far-reaching consequences on $\NN=3$ supergravity. This variant would use the maximum number of compensating vector multiplets coupled to the standard Weyl multiplet. Hence one may expect to use this variant for a construction of $\NN=3$ Poincar{\'e} supergravity from conformal supergravity without the need of any compensating multiplets. Hence such a construction of $\NN=3$ Poincar{\'e} supergravity would be completely off-shell in contrast to earlier constructions \cite{Hegde:2022wnb} which are on-shell because they rely on on-shell compensating vector multiplets.

In \cite{Hegde:2022wnb}, the authors developed higher-derivative invariants for $\mathcal{N}=3$ Poincar{\'e} supergravity in four dimensions by deriving the Poincaré supergravity action up to fourth-order in derivatives. In order to do this, the authors used the conformal supergravity action for the standard Weyl multiplet. The dilaton Weyl multiplet that we construct offers an alternative approach in formulating $\mathcal{N}=3$ Poincaré supergravity. In this alternate formulation, one would need to construct an $\NN=3$ conformal supergravity action based on the dilaton Weyl multiplet. The invariant action constructed out of a standard Weyl multiplet is fourth order in derivatives because the action has to be Weyl invariant. However, the dilaton Weyl multiplet has a dilaton field that has a non-trivial Weyl weight, and hence, in principle, one can have an invariant action constructed out of the dilaton Weyl multiplet that has terms containing more than 4-derivatives which would come with appropriate negative powers of the dilaton to maintain the Weyl invariance of the action. It will be intriguing to see whether one could construct off-shell invariants that involve more than four derivatives by employing the dilaton Weyl multiplet. For this purpose, one can use the action principle discussed in \cite{Butter:2019edc} to construct invariant actions with the dilaton Weyl multiplet. We reserve this investigation for future work.

\acknowledgments
We thank Chennai Mathematical Institute, IIT Ropar (and the organisers of NSM, 2024), ICTS Bengaluru and IISER Thiruvananthapuram for their hospitality during the course of this work. We thank Subramanya Hegde for useful discussions. {This research of MM was supported by appointment to the Young Scientist Training (YST) Program at the APCTP through the Science and Technology Promotion Fund and Lottery Fund of the Korean Government.} 
			
			\appendix
   \section{R-symmetry breaking} \label{decomposition}

Here, 
In table \eqref{break0}, we give the details of how fields of the $\NN=3$ standard Weyl multiplet, as well as vector multiplet in the $SU(3)$ representation, decompose into $SU(2)$ representation.

\begin{table}[h!]
	\centering
	\centering
	\begin{tabular}{ |p{4cm}|p{6cm}|}
		\hline
		$SU(3)$ irreps & $SU(2)$ irreps\\
			\hline
		\multicolumn{2}{|c|}{Fields of the standard Weyl Multiplet}\\
		\hline
$e_\mu^a(\bf{1})$&$e_\mu^a(\bf{1})$\\
  $\psi_\mu^I(\bf{3})$ & $\psi_\mu^i(\bf{2})$\;, $\psi_{\mu L} $$(\bf{1})$\\
  $V_\mu{}^I{}_J(\bf{8}) $ & $v_\mu{}^i{}_j(\bf{3})$\;, $Y_a^i(\bf{2})$\;, $\mathring{v}_\mu(\bf{1}).$\\ 
  $A_\mu (\bf{1}) $ & ${A}_\mu(\bf{1}) $\\
		$T_{ab}^{I}$  $(\bf{5})$& $T_{ab}^i$ $(\bf{2})$\;,\;$\mathring{T}_{ab}^+$  $(\bf{1})$\\
  $E_I(\bf{\bar{3}})$ & $E_i(\bf{{2}})$\;, $E$ (\bf{{1}})\\
  $D^I{}_J$ (\bf{8}) & $\Tilde{D
  }^i{}_j$(\bf{3})\;, $\mathring{D}^i(\bf{2})$\;, $\mathring{D}(\bf{1})$.\\
 $ \zeta^I$ (\bf{3})& $\zeta^i(\bf{2})$\;, $\mathring{\zeta}_L(\bf{1})$\\
 $\Lambda_L$(\bf{1})& $\Lambda_L$(\bf{1})\\
		$\chi_{IJ}$  $(\bf{\bar{6}})$&$\chi_{ij}$$(\bf{{3}})$\;, $\mathring{\chi}_i$$(\bf{{2}})$\;, $\mathring{\chi}_L (\bf{1})$.\\
		\hline
		\multicolumn{2}{|c|}{Fields of the vector Multiplet}\\
		\hline
$C_{\mu}$ $(\bf{1})$& $C_{\mu} $ $(\bf{1})$\\
$\xi_I  $$(\bf{\bar{3}})$& $\xi_i$$(\bf{{2}})$\;, $\xi(\bf{1})$\\
$\psi_I(\bf{\bar{3}}) $& $\psi_i (\bf{{2}})$\;, $\psi_R(\bf{1}) $\\
$\theta_L(\bf{1}) $ & $\theta_L(\bf{1}) $\\
 \hline
	\end{tabular}
	\caption{Decomposition of $SU(3)$ to $SU(2)$}
	\label{break0}	
\end{table}
The relations of the fields in the RHS of the table to those in the LHS of the table and are given in \eqref{1relations}.
\begin{align}\label{1relations}
&\psi_\mu^i = \psi_\mu^i, \quad \psi_\mu^3= \psi_{\mu L} \nonumber\\
&V_{\mu}{}^{i}{}_j + \frac{1}{2} \delta^i_j V_{\mu}{}^{3}{}_{3} = v_{\mu}{}^{i}{}_{j}, \quad -iV_{\mu}{}^{3}{}_{3} = v_{\mu}, \quad V_{a}{}^i{}_{3} - \frac{1}{2} u(\psi_a)^i = Y_a^i \nonumber\\
&T^i_{ab} = T^i_{ab}, \quad T^3_{ab} = \mathring{T}^{+}_{ab} \nonumber\\
&E_i = E_i, \quad E_3 =E\nonumber\\
&D^{i}{}_{j} - \frac{1}{2}\delta^i_j D^{3}{}_{3} = \tilde{D}^i{}_j, \quad D^3{}_3 = \mathring{D}, \quad D^{i}{}_3 = \mathring{D}^i\nonumber \\
&\zeta_i = \zeta_i, \quad \zeta_3 = \mathring{\zeta}_R \nonumber \\
& \chi_{ij}= \chi_{ij}, \quad \chi_{i3}= \mathring{\chi}_i, \quad \chi_{33} = \mathring{\chi}_L \nonumber \\
&\xi_i = \xi_i, \quad \xi_3 = \xi \nonumber \\
& \psi_i = \psi_i, \quad \psi_3 = \psi_R
\end{align}

In the following as well as in the main sections only $\tilde{D}^i{}_j$ appears. Hence, we conveniently drop the ``tilde'' and call it $D^i{}_j$.

The reality property of some of the decomposed fields is induced from the reality property
of the original fields and is given as below:
\begin{align}
(v_\mu{}^i{}_j)^*&=-v_\mu{}^j{}_i\nonumber\\
(v_\mu)^*&=v_\mu\nonumber\\
(D^i{}_j)^*&=D^j{}_i \nonumber\\
(\mathring{D})^*&=\mathring{D}
\end{align}

			\bibliography{references}
			\bibliographystyle{jhep}

\end{document}